\documentclass[%
reprint,
 superscriptaddress,
 amsmath,amssymb,
 aps,
 nofootinbib,
 prb, preprintnumbers
]{revtex4-2}

\usepackage{textgreek}
\usepackage{multirow}
\usepackage{bigdelim}
\usepackage[dvipdfmx]{graphicx}
\usepackage{color}
\usepackage{dcolumn}
\usepackage{physics}
\usepackage{mathtools}
\usepackage{bm}
\usepackage{array}
\usepackage{comment}
\usepackage{cancel}
\usepackage{here}
\usepackage[usenames,dvipsnames]{xcolor}
\usepackage{mathrsfs}
\usepackage{amsmath,amsthm}
\usepackage{enumerate}
\usepackage{empheq} 
\theoremstyle{definition}

\usepackage{xcolor} 
\definecolor{RoyalBlue}{rgb}{0.255, 0.412, 0.882}
\definecolor{DeepSkyBlue}{rgb}{0.0, 0.749, 1.0}
\definecolor{Crimson}{rgb}{0.862, 0.078, 0.235}
\definecolor{ForestGreen}{rgb}{0.133, 0.545, 0.133}
\definecolor{OrangeRed}{rgb}{1.0, 0.271, 0.0}
\definecolor{Orchid}{rgb}{0.855, 0.439, 0.839}
\definecolor{Sienna}{rgb}{0.627, 0.322, 0.176}
\definecolor{Goldenrod}{rgb}{0.855, 0.647, 0.125}
\definecolor{CadetBlue}{rgb}{0.372, 0.619, 0.627}
\definecolor{CornflowerBlue}{rgb}{0.392, 0.584, 0.929}
\definecolor{RebeccaPurple}{rgb}{0.4, 0.2, 0.6}
\definecolor{Salmon}{rgb}{0.980, 0.502, 0.447}
\definecolor{HotPink}{rgb}{1.0, 0.412, 0.706}
\definecolor{Chocolate}{rgb}{0.824, 0.412, 0.118}
\definecolor{SteelBlue}{rgb}{0.275, 0.510, 0.706}
\definecolor{FireBrick}{rgb}{0.698, 0.133, 0.133}
\definecolor{bondiblue}{rgb}{0.0, 0.58, 0.71}
\definecolor{celestialblue}{rgb}{0.29, 0.59, 0.82}
\definecolor{coolblack}{rgb}{0.0, 0.18, 0.39}
\definecolor{frenchblue}{rgb}{0.0, 0.45, 0.73}
\definecolor{lapislazuli}{rgb}{0.15, 0.38, 0.61}
\definecolor{mediumpersianblue}{rgb}{0.0, 0.4, 0.65}
\definecolor{darkpowderblue}{rgb}{0.0, 0.2, 0.6}
\definecolor{darkcandyapplered}{rgb}{0.64, 0.0, 0.0}
\definecolor{darkscarlet}{rgb}{0.34, 0.01, 0.1}
\definecolor{falured}{rgb}{0.5, 0.09, 0.09}
\definecolor{darkcyan}{rgb}{0.0, 0.55, 0.55}

\newcommand{\ri}{\mathrm{i}}
\def\beq{\begin{equation}}
\def\eeq{\end{equation}}
\def\bp{\begin{pmatrix}}
\def\ep{\end{pmatrix}}

\usepackage[pdftex,colorlinks,urlcolor=darkcyan,citecolor=blue,linkcolor=blue]{hyperref}

\begin{document}

\preprint{USTC-ICTS/PCFT-24-19}

\author{Yunfeng Jiang}
\affiliation{School of physics \& Shing-Tung Yau Center, Southeast University, Nanjing  211189, P. R. China}
\affiliation{Peng Huanwu Center for Fundamental Theory, Hefei, Anhui 230026, China}

\author{Yuan Miao}
\affiliation{Kavli Institute for the Physics and Mathematics of the Universe (WPI), UTIAS, The University of Tokyo, Kashiwa, Chiba 277-8583, Japan}

\date{\today}

\title{Spectrum-preserving Deformations of Integrable Spin Chains with Open Boundaries}
\begin{abstract}

We discover a family of local deformations that leave part of the spectrum intact for strongly interacting and exactly solvable quantum many-body systems. Since the deformation preserves the Bethe Ansatz equations~(BAE), it is dubbed the iso-BAE flow. Although all theories on the flow share the same BAE, the spectra are different. Part of the spectrum remains intact along the whole flow. Such states are protected by an emergent symmetry. The remaining parts of the spectrum change continuously along the flow and are doubly degenerate for even length spin chains. For odd length chains, the deformed spectrum also comprises doubly degenerate pairs apart from the sector with magnon number $(L+1)/2$, where $L$ is the length of the spin chain. We discuss the iso-BAE flow for the ${\rm XXX}_{1/2}$ model in detail and show that the iso-BAE flows exist for more general models including $q$-deformed XXZ as well as higher spin ${\rm XXX}_{s}$ spin chains.

\end{abstract}

\maketitle

\section{Introduction}
\label{sec:intro}

In quantum mechanics, we frequently encounter the problem of finding the spectrum of the Hamiltonian $\mathbf{H}_{\alpha}=\mathbf{H}_0+\mathbf{V}_{\alpha}$, where $\mathbf{H}_0$ is solvable and $\mathbf{V}_{\alpha}$ is a local deformation (\emph{i.e.} $\mathbf{V}_{\alpha}$ is a sum of local interactions) with $\alpha$ being a continuous parameter. The spectrum of $\mathbf{H}_{\alpha}$ can be found by perturbation theory~\cite{Landau:1991wop, Cohen-Tannoudji:101367} or other approximation methods such as Hamiltonian truncation method~\cite{Yurov:1989yu, Yurov:1991my, James:2017cpc}. Generically, the spectra of $\mathbf{H}_0$ and $\mathbf{H}_{\alpha}$ are different and the eigenvalues of $\mathbf{H}_{\alpha}$ depend on the parameter $\alpha$. In some situations, $\mathbf{H}_{\alpha}$ is related to $\mathbf{H}_0$ by a unitary transformation, resulting in identical spectra. An interesting general question is: are there deformations $\mathbf{V}_{\alpha}$ such that $\mathbf{H}_{\alpha}$ and $\mathbf{H}_0$ share \emph{part of the spectrum}. Namely, part of the eigenvalues remain intact against the perturbation while the rest are deformed. Such deformations, if exist, may reveal the existence of hidden or emergent symmetry which protect the invariant spectra. How to construct such deformations for a generic quantum many-body system is unclear. In this work, we aim to offer insights on a strongly interacting and exactly solvable set-up. In particular, we focus on quantum integrable models and provide a class of such examples.

We will consider quantum integrable spin chains which can be solved by Bethe ansatz \cite{Bethe_1931, Gaudin_2014, Korepin_1993}. 
Integrability provides us powerful tools to analyze the spectrum and offers useful guidance for the construction of deformations that preserve part of the spectrum. As it turns out, such a deformation is linked to another phenomenon in Bethe ansatz, which has been hitherto unexplored (see \cite{Nepomechie:2019gqt} for an exception). We find that the deformed models share the same Bethe ansatz equations (BAE), hence dubbed an iso-BAE flow. The spectra of models on the flow are closely related.  More precisely, eigenstates of the Hamiltonian $\mathbf{H}_{\alpha}$ fall into two categories, referred to as type-I and type-II respectively. The eigenvalues of type-I states remain invariant along the flow while those of the type-II states change with $\alpha$.\par

We will analyze the flow by two complementary approaches. In the first approach, we construct an operator $\mathcal{Q}_{\alpha}$ and show that it commutes with the Hamiltonian $[\mathbf{H}_{\alpha},\mathcal{Q}_{\alpha}]=0$ and is nilpotent {(for even length)} $\mathcal{Q}_{\alpha}^2=0$. It then follows that, once acting $\mathcal{Q}_{\alpha}$ on a given eigenstate $\mathcal{Q}_{\alpha}|\psi\rangle$ there are only two possibilities: either the result is vanishing or we obtain another eigenstate which is degenerate with $|\psi\rangle$, which correspond to type-I and type-II states, respectively.
We then study more closely the eigenvalues by a careful scrutiny of BAE. To perform this analysis, we use the rational $Q$-system \cite{Marboe:2016yyn, Granet:2019knz, Bajnok:2019zub, Nepomechie:2019gqt, Hou:2023jkr, Hou:2023ndn}, which gives the complete physical spectrum. From the $Q$-system, we find that the results are also in accord with the classification of states mentioned before.\par

We first perform the analysis for the XXX$_{1/2}$ type spin chain for simplicity. Then we will show that the construction can be generalized to the $q$-deformed and higher spin models. While each generalization has certain new features, the main characteristic, namely the undeformed BAE and the classification of type-I and type-II states, remains the same as in the XXX$_{1/2}$ case.

\section{The model and Iso-BAE flow}
\label{sec:openchains}
We start with the spin-$1/2$ open XXX (${\rm XXX}_{1/2}$) spin chain \cite{Alcaraz1987, Sklyanin:1988yz} with free boundary conditions
\begin{align}
\mathbf{H}_0=\sum_{n=1}^{L-1} \left( \vec{\sigma}_n \cdot \vec{\sigma}_{n+1} - 1 \right)\,
\label{eq:H0XXX}
\end{align}
where $\vec{\sigma}_n=(\sigma_n^x,\sigma_n^y,\sigma_n^z)$ are the Pauli matrices acting on site $n$. This model is SU(2) invariant and the eigenstates form SU(2) multiplets. Now we deform the model by coupling it to two parallel boundary magnetic fields and define
\begin{align}
\label{eq:Hflow}
\mathbf{H}_{\alpha}=\mathbf{H}_0+\mathbf{V}_\alpha,\qquad \mathbf{V}_{\alpha}=\frac{\sigma^z_L - 1}{\alpha} - \frac{\sigma^z_1 - 1}{\alpha-1}\,.
\end{align}
where $\alpha=\infty$ correspond to the undeformed case. $\mathbf{H}_{\alpha}$ is a special case of the following model~\cite{Sklyanin:1988yz, Nepomechie:2019gqt}
\beq 
    \mathbf{H}_{\alpha,\beta}= \sum_{n=1}^{L-1} \left( \vec{\sigma}_n \cdot \vec{\sigma}_{n+1} - 1 \right) + \frac{\sigma^z_L - 1}{\alpha} - \frac{\sigma^z_1 - 1 }{\beta}.
\label{eq:Hamiltonian}
\eeq 
at $\beta =\alpha-1$. We will also use the notation
\beq 
    h_L = -\frac{1}{\alpha-1} , \quad h_R = \frac{1}{\alpha} ,
\eeq
to parametrize the boundary magnetic fields when convenient.

$\mathbf{H}_{\alpha,\beta}$ is integrable for any boundary magnetic fields $\alpha, \beta \in \mathbb{C}$ and can be solved by Bethe ansatz \cite{Alcaraz1987, Gaudin_2014, Sklyanin:1988yz, Korepin_1993}. 

\subsection{Integrability}
Following standard Bethe ansatz approach, the eigenstates of \eqref{eq:Hamiltonian} can be parametrized by the Bethe roots $|\{ u_m \}_{m=1}^M \rangle$, where $M$ is the number of magnons (down spins). The Bethe roots are the solutions of the following BAE,
\beq 
    \frac{g(u_m-\ri/2)}{f(u_m +\ri/2 )}\left( \frac{u_m + \ri/2 }{u_m - \ri /2} \right)^{2L} \prod_{k\neq m}^M S(u_m, u_k) = 1 ,
    \label{eq:BAE}
\eeq 
where
\beq 
    S(u_m, u_k) = \frac{u_m - u_n - \ri }{u_m - u_n + \ri} \frac{u_m + u_n - \ri }{u_m + u_n + \ri}\,,
    \label{eq:scatteringphase}
\eeq 
and $f(u)$ and $g(u)$ are related to boundary magnetic fields
\beq 
    f(u) = g(-u) = (u- \ri \alpha) (u+\ri \beta) .
\eeq 
The corresponding energy for an eigenstate $|\{u_m\}\rangle$ is given by 
\beq 
    E_M=-\sum_{m=1}^M\frac{2}{u_k^2+1/4}\,.
    \label{eq:Eeigenvalue}
\eeq 
An interesting observation \cite{Nepomechie:2019gqt, Aubin:2015afa} is that when $\beta = \alpha - 1$, we have $f(u +\ri/2 ) = g(u -\ri/2 )$, and they cancel out in \eqref{eq:BAE}, leading to
\beq 
    \left( \frac{u_m + \ri/2 }{u_m - \ri /2} \right)^{2L} \prod_{k\neq m}^M S(u_m, u_k) = 1 .
    \label{eq:BAEsimple}
\eeq
Naively \eqref{eq:BAEsimple} can be identified with the BAE of \eqref{eq:Hflow}.~\footnote{When $\alpha = -\beta = 1$, the BAE for $(L-1)$ sites coincide with the BAE for $L$ site with $\beta = \alpha-1$. We explain the details in Sec. \ref{app:alpha1limit} of SM.} However, one immediately sees a problem because the BAE \eqref{eq:BAEsimple} does not depend on the boundary parameter $\alpha$ while the Hamiltonian itself does ! In Sec.~\ref{sec:BAE}, we will see that this is because \eqref{eq:BAEsimple} only describes part of the spectrum. Since the models \eqref{eq:Hflow} share the same BAE, we call the models with different $\alpha$ an iso-BAE flow.

\subsection{Type-I and type-II states}
\label{subsec:typeIandII}

We will now show that for the model $\mathbf{H}_{\alpha}$, there exists an operator $\mathcal{Q}_{\alpha}$ commuting with the Hamiltonian, \emph{i.e.}
\begin{align}
\label{eq:propQ}
[\mathcal{Q}_{\alpha},\mathbf{H}_{\alpha}]=0.
\end{align}
Explicitly, the operator is given by
\begin{align}
\mathcal{Q}_{\alpha}=\sum_{M=0}^{\lfloor L/2 \rfloor}\Pi_{L-M+1}\hat{\mathfrak{q}}_{\alpha}^{L-2M+1}\Pi_M\,.
\end{align}
Here $\Pi_M$ is the projector which projects to the $M$-magnon sector of the model with respect to the ferromagnetic state $|\uparrow \uparrow \cdots \uparrow \rangle$ and $\hat{\mathfrak{q}}_{\alpha}$ reads
\beq     \hat{\mathfrak{q}}_{\alpha}= - 2 \left( \alpha \mathbf{S}^- + \sum_{m<n} \sigma^-_m \sigma^z_n \right)\,,
    \label{eq:spinhalfq}
\eeq 
with $\mathbf{S}^-= \sigma_1^-+\ldots\sigma_L^-$ being the global spin lowering operator. From the construction, one can see that acting $\mathcal{Q}_{\alpha}$ on an $M$-magnon state leads to an $(L-M+1)$-magnon state. 

One special case occur for odd $L$ and $M=(L+1)/2$ where $\Pi_{L-M+1}\hat{\mathfrak{q}}_{\alpha}^{L-2M+1}\Pi_M=\Pi_M$ becomes a projector. {Using the property of the projectors, we have
\beq 
    \mathcal{Q}_\alpha^2 = \begin{cases} \Pi_{(L+1)/2} , & \, L \, \mathrm{mod} \, 2 = 1 , \\
    0 , & \,  L \, \mathrm{mod} \, 2 = 0 . \end{cases} 
\eeq 
Hence, operator $\mathcal{Q}_\alpha$ is nilpotent for even length.} 

Commutativity of \eqref{eq:propQ} and $\mathbf{H}_{\alpha}$ is proven in Sec.~\ref{app:ABAXXX} of SM. We would like to stress that although the operator $\mathcal{Q}_{\alpha}$ commutes with the Hamiltonian, it is not one of the higher conserved charges which can be generated from the transfer matrix, \emph{i.e.} it does not belong to the Bethe algebra. The existence of $\mathcal{Q}_{\alpha}$ reflects a novel emergent symmetry of the iso-BAE flow.

Due to the property \eqref{eq:propQ}, for any given eigenstate $|\psi_{\alpha}\rangle$ of the model, $\mathcal{Q}_{\alpha}|\psi_{\alpha}\rangle$ is either vanishing or is another eigenstate with the same eigenvalue, but with a different magnon number. It is straightforward to check that both cases exist. For $M\le\lfloor L/2\rfloor$, we call the eigenstates that are annihilated by $\mathcal{Q}_{\alpha}$ type-I states, the rest of the eigenstates are type-II states.

\begin{figure}[h!]
    \centering
    \includegraphics[width=\linewidth]{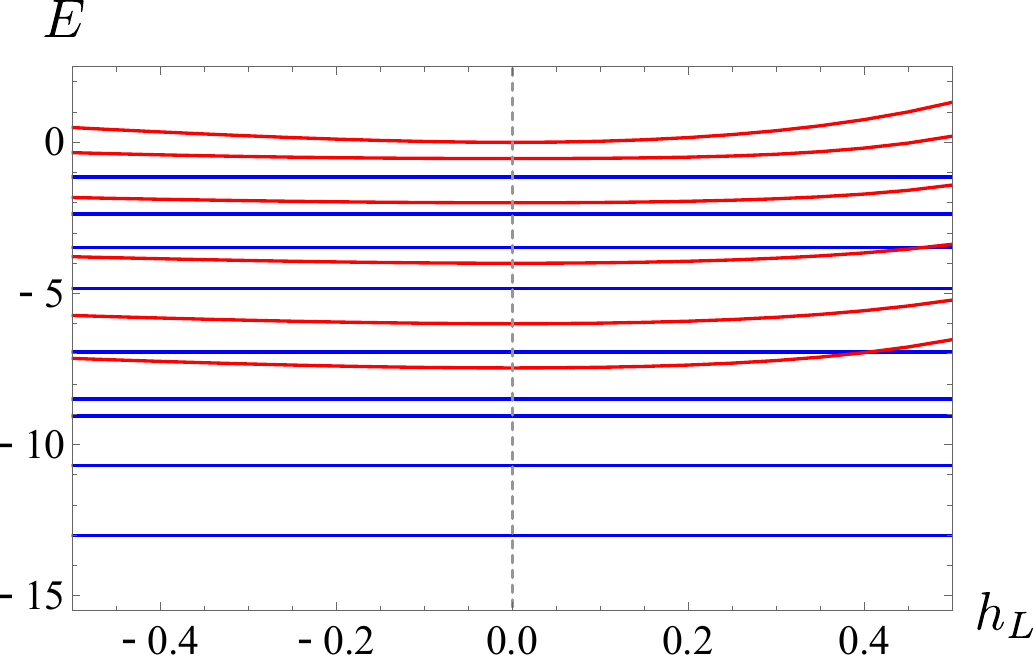}
    \caption{The energy spectrum of the iso-BAE system as a function of $h_L = \frac{1}{\alpha}$ for $L=6$ and $M=2$. The type-I states are labelled in blue while the type-II ones in red. The number of type-I states is $\bp 6 \\ 2 \ep - \bp 6 \\ 1 \ep = 9$, while the number of type-II states is $\bp 6 \\ 1 \ep = 6$.}
    \label{fig:energy_L6_M2}
\end{figure}

The eigenvalues of the type-I states do not depend on $\alpha$ and they remain invariant along the flow. On the other hand, the eigenvalues of type-II states change with $\alpha$, as are shown in Fig.~\ref{fig:energy_L6_M2}. In this sense, we can say that the type-I states are `protected' by the emergent symmetry $\mathcal{Q}_{\alpha}$. This observation is based on extensive numerical checks, while it would be desirable to have a more rigorous proof. Yet another compelling observation from numerical results is that the type-I states actually span the kernel of $\mathcal{Q}_{\alpha}$.

\subsection{Structure of the Hilbert space}
\label{subsec:structureHilbert}

We now discuss the structure of the Hilbert space. 
\paragraph{The undeformed model}
For the undeformed model (corresponding to $\alpha=\infty$), which is SU(2) invariant, eigenstates form SU(2) multiplets. The highest weight states correspond to physical solutions of the BAE \eqref{eq:BAEsimple}. Moreover, the operator $\hat{\mathfrak{q}}_{\alpha}$ is proportional to the global spin lower operator $\mathbf{S}^-$. Therefore, all the highest-weight states are type-I states, while all the descendants are type-II. In the $M$-magnon sector, there are $\mathcal{N}_{\text{I}}(L,M)= \bp L \\ M \ep - \bp L \\ M-1 \ep$ physical solutions. Each solution corresponds to a spin-$(\frac{L}{2}-M)$ representation with dimension $(L-2M+1)$. 

\begin{figure}[h!]
\centering
\includegraphics[scale=0.5]{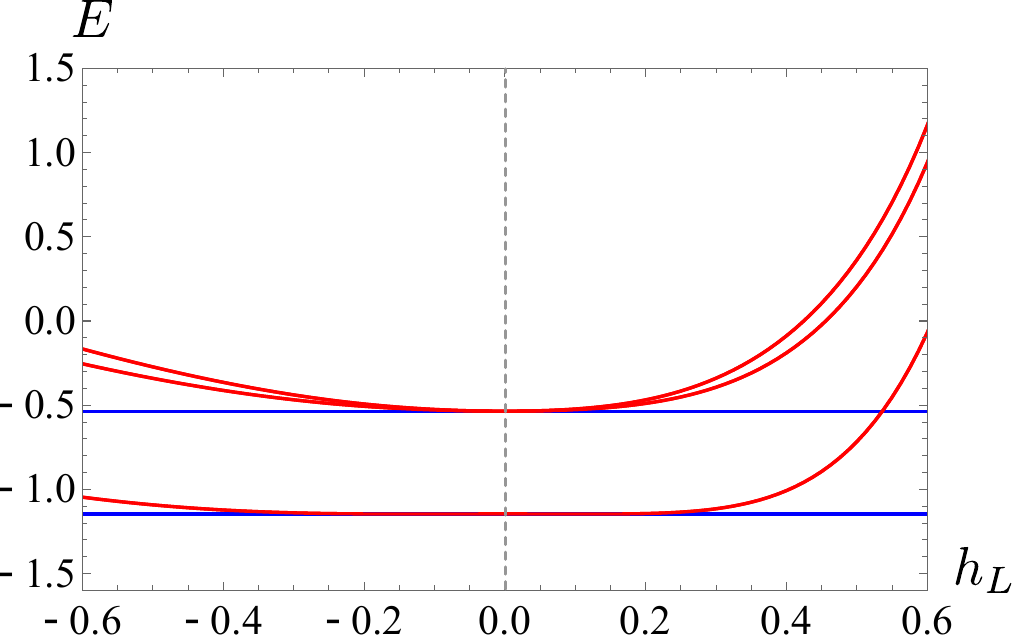}
\caption{Lifting of the degeneracies for the spectrum for finite $\alpha$. The blue lines correspond to the energy eigenvalues for the type-I states, while the red lines are the type-II states. In this figure, the two type-I states are $M=2$ and $M=1$ for system size $L=6$ from bottom to top respectively. The type-II states are doubly degenerate in this case. Here we present eigenstates with different magnetization sectors, which are different from the ones in Fig. \ref{fig:energy_L6_M2}.}
\label{fig:split}
\end{figure}

\paragraph{The deformed model}
For finite $\alpha$, the SU(2) symmetry is broken by $\mathbf{V}_{\alpha}$. The highest weight states at $\alpha=\infty$ become type-I states at finite $\alpha$. The descendant states at $\alpha=\infty$ become type-II states and their energy eigenvalues change with $\alpha$. In other words, the degeneracies are lifted, this is shown in Fig.~\ref{fig:split}.

While lifting degeneracies after certain deformation is a common feature, what is special for the iso-BAE flow is that the type-II states are organized by the emergent symmetry $\mathcal{Q}_{\alpha}$ and have residual degeneracies. For even $L$, the Hilbert space $\mathcal{H}$ has the following decomposition
\begin{align}
\mathcal{H}_L=\bigoplus_{M=0}^{L/2}\{|\psi_M\rangle^{\text{I}}\}\oplus\{|\psi_M\rangle^{\text{II}},\mathcal{Q}_{\alpha}|\psi_M\rangle^{\text{II}}\}
\label{eq:evenHilbert}
\end{align}
where $|\psi_M\rangle^{\text{I}}$ and $|\psi_M\rangle^{\text{II}}$ stand for type-I and type-II states with $M$ magnons.
For odd $L$, we have the following decomposition
\beq 
\begin{split}
\mathcal{H}_L=\,\bigoplus_{M=0}^{(L-1)/2} & \{|\psi_M\rangle^{\text{I}}\} \oplus\{|\psi_M\rangle^{\text{II}},\mathcal{Q}_{\alpha} |\psi_M\rangle^{\text{II}}\}\\
&\,\oplus\{|\psi_{(L+1)/2}\rangle^{\text{II}}\}
\end{split}
\label{eq:oddHilbert}
\eeq 
The sector $M=(L+1)/2$ is special because in this sector $\mathcal{Q}_{\alpha}=\Pi_{(L+1)/2}$ and $\mathcal{Q}_{\alpha}|\psi_{(L+1)/2}\rangle^{\text{II}}=|\psi_{(L+1)/2}\rangle^{\text{II}}$.\par

We will show that in the $M$-magnon sector ($M\leq \lfloor \frac{L}{2} \rfloor$), the number of type-I and type-II states are $\mathcal{N}_{\text{I}}(L,M)$ and $\mathcal{N}_{\text{II}}(L,M)={L\choose M-1}$ respectively. This can be seen clearly in Fig.~ \ref{fig:energy_L6_M2} for the iso-BAE systems with 6 sites.

For $M\leq \lfloor \frac{L}{2} \rfloor$, the state $\mathcal{Q}_{\alpha}|\psi_M\rangle^{\text{II}}$ is an $M'=L-M+1$ eigenstate with the same eigenvalue. Therefore, there is an effective `equator' at $\tfrac{1}{2}(L+1)$. For $M > \lfloor L/2\rfloor$, there are no type-I eigenstates and the number of type-II eigenstates is ${ L \choose M }$. It is straightforward to check that 
\begin{align*}
\sum_{M=0}^{\lfloor \frac{L}{2}\rfloor}(\mathcal{N}_{\text{I}}(L,M)+\mathcal{N}_{\text{II}}(L,N))+\sum_{M=\lfloor \frac{L}{2}\rfloor+1}^L{ L \choose M }=2^L\,.
\end{align*}
We thus conclude that the type-I and type-II states for different magnon numbers constitute the complete Hilbert space.

A natural question is whether the ground state is type-I or type-II ? It turns out that this depends on the value of $\alpha$, and we have two phase transition points at $\alpha=0$ and $\alpha=1$. A detailed discussion on the phases of the ground state is given in Sec. \ref{app:phases} of SM.

\section{BAE, \texorpdfstring{$TQ$}{TQ} and \texorpdfstring{$QQ$}{QQ} analysis}
\label{sec:BAE}
We will now analyze the spectrum of the iso-BAE flow from the perspective of Bethe roots and show that we can reach the same result as in the previous sections. As alluded before, the solutions to BAE \eqref{eq:BAEsimple} only describe part of the spectrum. To find out the `missing' solutions, we need to consider Baxter's $TQ$-relation
\beq 
\begin{split}
    - u T_{1/2}(u) Q(u) =&\,  u^+ T_0^+ (u) g^- (u) Q^{2-}  (u)\\
    & +  u^- T_0^-  (u) f^+ (u) Q^{2+}  (u) ,
\end{split}
\label{eq:TQrel}
\eeq
where we use the shorthand notation $f^{n \pm} = f(u \pm n \ri /2)$. Here $T_{1/2}(u)$ is the eigenvalue of the transfer matrix (cf. Sec. \ref{app:ABAXXX} of SM), $T_0 (u) = u^{2L}$ and
 $Q(u)$ is the Baxter's $Q$-polynomial
\beq 
    Q(u) = \prod_{m=1}^M (u - u_m)(u + u_m) 
\eeq 
whose zeros are the Bethe roots. For generic boundary magnetic fields $\alpha$ and $\beta$, taking $u=u_m$ in \eqref{eq:TQrel} gives BAE in \eqref{eq:BAE}. For $\beta =\alpha-1$, the $TQ$-relation \eqref{eq:TQrel} can be written as
\begin{align}
\label{eq:TQrelation}
-u\,T_{1/2}Q=\big( u^2+(\alpha-\tfrac{1}{2})^2 \big)\left(u^+T_0^+ Q^{2-}+ u^-T_0^- Q^{2+}\right)
\end{align}
For $u=\pm\mathrm{i}(\alpha-\tfrac{1}{2})$, the right-hand side vanishes. If $Q(u)$ contains the roots $u=v_M=\pm\mathrm{i}(\alpha-\tfrac{1}{2})$, the $TQ$-relation \eqref{eq:TQrelation} can still be satisfied. Let us denote the rest of the Bethe roots by $\{v_j\}_{j=1}^{M-1}$, it follows that $v_j$ satisfy the following reduced BAE
\beq 
    \frac{\tilde{g}(v_m-\ri/2)}{\tilde{f}(v_m +\ri/2 )}\left( \frac{v_m + \ri/2 }{v_m - \ri /2} \right)^{2L} \prod_{k\neq m}^{M-1} S(v_m, v_k) = 1 ,
    \label{eq:reducedBAE}
\eeq 
where $S(u,v)$ is the same as \eqref{eq:scatteringphase} and the polynomials $\tilde{f} (u)$ and $\tilde{g} (u)$ are
\beq 
    \tilde{f} (u ) = \tilde{g} (-u ) = (u +\ri \alpha ) (u - \ri (\alpha-1) ) .
\eeq 
This implies that the set of $(M-1)$ Bethe roots $\{ v_m \}_{m=1}^{M-1}$ solves the BAE with $\tilde{\alpha} = - \alpha $ and $\tilde{\beta} = - \beta = -(\alpha-1)$, \emph{i.e.} the BAE for the Hamiltonian with opposite boundary magnetic fields. To conclude, we have the following two types of solutions 
\begin{itemize}
\item Type-I. $\{u_m\}$, $m=1,\ldots,M$ which satisfies \eqref{eq:BAEsimple}; These Bethe roots do not depend on the boundary magnetic fields and the corresponding eigenvalues remain invariant under the iso-BAE flow;
\item Type-II. $\{\ri (\alpha-\tfrac{1}{2})\}\cup\{v_1,\ldots,v_m\}$, $m=1,\ldots,M-1$ where $\{v_m\}$ satisfy the reduced BAE \eqref{eq:reducedBAE}. These Bethe roots depend on the boundary magnetic field in a non-trivial manner and the corresponding eigenvalues flow with $\alpha$.
\end{itemize}
Obviously the two types of solutions correspond to the two kinds of states that we discussed in the previous section.
Let us comment that, in order to find all the physical solutions of the BAE, we use the rational $Q$-system approach, which is reviewed in Sec. \ref{app:QsystemXXX} of SM. From the solution of rational $Q$-system, one can check explicitly that the number of solutions for type-I and type-II states are indeed $\mathcal{N}_{\text{I}}(L,M)$ and $\mathcal{N}_{\text{II}}(L,M)$ respectively.

\section{Generalizations}
\label{sec:generalizations}
So far we have discussed the iso-BAE flow for the spin-$1/2$ isotropic XXX spin chain. In this section, we consider two natural generalizations. One is the $q$-deformed spin-$1/2$ XXZ model and the other is the higher spin XXX$_s$ model.

\subsection{XXZ model}
\label{subsec:XXZ}

For the XXZ model, we construct the following iso-BAE flow
\beq 
    \mathbf{H}^{\rm XXZ}_{\alpha} = \mathbf{H}^{\rm XXZ}_{0} + \mathbf{V}^{\rm XXZ}_{\alpha} . 
\label{eq:isoXXZ}
\eeq 
where $\mathbf{H}_0^{\text{XXZ}}$ is the $U_q (\mathfrak{sl}_2)$-invariant XXZ model~\cite{Pasquier:1989kd},
\beq 
\begin{split}
    \mathbf{H}^{\rm XXZ}_0 = & \sum_{n=1}^{L-1} \left[ \sigma^x_n \sigma^x_{n+1} + \sigma^y_n \sigma^y_{n+1} \right. \\
    & \left. + \Delta (\sigma^z_n \sigma^z_{n+1} - 1 ) \right]
     + \sinh \eta (\sigma^z_L - \sigma^z_1) ,
\end{split}
\eeq 
with $\Delta=\tfrac{q+q^{-1}}{2}=\cosh\eta$ being the anisotropy. 

The deformation term is given by
\beq 
\begin{split}
    \mathbf{V}^{\rm XXZ}_{\alpha} & = \sinh \eta \left[ \big( \frac{1}{\tanh (\alpha \eta)} -1 \big) (\sigma^z_L - 1) \right. \\
    & \left. - \big( \frac{1}{\tanh ((\alpha-1)\eta)} -1 \big) (\sigma^z_1 - 1) \right]
\end{split}
\eeq 
where $\eta \neq 0$ is either real or imaginary, corresponding to two regimes $|\Delta| > 1$ and $0<|\Delta| < 1$.

The model \eqref{eq:isoXXZ} can be solved by Bethe ansatz. The invariant part of the spectrum is described by the following BAE
\beq 
    \left( \frac{\sinh(u_m+\eta/2)}{\sinh(u_m-\eta/2)} \right)^{2L} \prod_{k\neq m}^M S^{\rm XXZ} (u_m, u_k) = 1 .
\eeq 
More details can be found in Sec. \ref{app:QsystemXXZ} of SM.

The structure of Hilbert space for the XXZ iso-BAE systems is similar to the XXX case. Here we focus on the model with generic $q$. The reason is that at root of unity, the representation theory of the quantum group $U_q (\mathfrak{sl}_2)$ is different~\cite{Pasquier:1989kd, Miao:2020irt}, which leads to additional degeneracies in the spectrum~\cite{Aubin:2015afa, Azat:2016pxy}. As a result, we need to treat the iso-BAE flow with more care, dealing with additional degeneracies at finite $\alpha$. 

We have type-I states that are invariant under the change of the boundary fields, as well as the type-II states that do vary. Similarly, we find that the type-I states span the kernel of the following operator,
\beq 
\mathcal{Q}_{\alpha} (q) =\sum_{M=0}^{\lfloor L/2 \rfloor}\Pi_{L-M+1}\hat{\mathfrak{q}}_{\alpha}^{L-2M+1} (q) \Pi_M\,.
\eeq 
Here $\Pi_M$ is the same projector which projects to the $M$-magnon sector of the model and $\hat{\mathfrak{q}}_{\alpha} (q)$ is the $q$-deformed version of $\hat{\mathfrak{q}}_{\alpha}$ defined in \eqref{eq:spinhalfq}
\beq 
\begin{split}
    \hat{\mathfrak{q}}_{\alpha} (q) & = \sum_{m=1}^L  \left( - q^{-\alpha-1} \prod_{n=m+1}^L q^{-\sigma^z_n} \right. \\
    & \left. + q^{\alpha-1} \prod_{n=m+1}^L q^{\sigma^z_n} \right) \sigma^-_m\,.
\end{split}
\eeq 

The Hilbert spaces for even and odd lengths of the chain are decomposed precisely into the same way in \eqref{eq:evenHilbert} and \eqref{eq:oddHilbert}, where the same double degeneracies for type-II states appear.

\subsection{Higher spin \texorpdfstring{$\mathrm{XXX}_s$}{XXXs} model}
\label{subsec:higherspinXXXs}

Another generalization of the spin-$1/2$ XXX model is the higher-spin XXX$_s$ model where the dimension of local Hilbert space is $2s+1$ with $2s \in \mathbb{N}$. \paragraph{The spin-1 model} One important example of higher spin models is the spin-1 Takhtajan-Babujian model \cite{Takhtajan:1982jeo, Babujian:1983ae}, \emph{i.e.} the $\mathrm{XXX}_1$ model.

For the $\mathrm{XXX}_1$ model, we can construct the following iso-BAE flow
\beq 
    \mathbf{H}^{(1)}_{\alpha} = \mathbf{H}^{(1)}_0 + \mathbf{V}_{\alpha}^{(1)} ,  
\eeq 
where 
\beq 
    \mathbf{H}^{(1)}_0 = \sum_{n=1}^{L-1} \left[ \vec{\mathbf{S}}_n \cdot \vec{\mathbf{S}}_{n+1} - \big( \vec{\mathbf{S}}_n \cdot \vec{\mathbf{S}}_{n+1}  \big)^2  \right] ,
\eeq 
and $\mathbf{V}^{(1)}_{\alpha}$ is defined in \eqref{eq:spin1Valpha}. The spin-1 operators are defined in \eqref{eq:spin1operators}.

The model $\mathbf{H}^{(1)}_0$ at $\alpha=\infty$ is SU(2) invariant and the eigenstates are again organized as the highest-weight states of the global SU(2) symmetry and their descendants.
At finite $\alpha$, the Hilbert space is decomposed as
\begin{align}
\mathcal{H}_{2L}^{(1)} =\bigoplus_{M=0}^{L}\{|\psi_M\rangle^{\text{I}}\}\oplus\{|\psi_M\rangle^{\text{II}},\mathcal{Q}_{\alpha}|\psi_M\rangle^{\text{II}}\} ,
\end{align}
where the maximal number of magnons becomes $2L$ and 
\beq 
\mathcal{Q}_{\alpha}^{(1)} =\sum_{M=0}^{L}\Pi_{2L-M+1}(\hat{\mathfrak{q}}^{(1)}_{\alpha})^{2L-2M+1} (q) \Pi_M\,,
\eeq 
\beq \hat{\mathfrak{q}}_{\alpha}^{(1)}  = -2 \alpha \sum_{m=1}^L \mathbf{S}^-_m - 4 \sum_{m<n} \mathbf{S}^-_m \mathbf{S}^z_n - \sum_{m=1}^L \{ \mathbf{S}^-_m , \mathbf{S}^z_m \} .
\eeq

\paragraph{Spin-$s$ model} Generalizations to higher $s$ is straightforward. The undeformed model $\mathbf{H}_0^{(s)}$ is taken as the SU(2) invariant spin-$s$ chain with free boundary condition. The corresponding deformation $\mathbf{V}_{\alpha}^{(s)}$ can be found in SM. For spin-$s$ models, the emergent symmetry reads
\beq 
\mathcal{Q}_{\alpha}^{(s)} =\sum_{M=0}^{\lfloor sL \rfloor}\Pi_{2 s L-M+1}(\hat{\mathfrak{q}}^{(s)}_{\alpha})^{2 s L-2M+1} \Pi_M\,,
\eeq 
where
\beq 
\hat{\mathfrak{q}}_{\alpha}^{(s)}  = -2 \alpha \sum_{m=1}^L \mathbf{S}^-_m - 4 \sum_{m<n} \mathbf{S}^-_m \mathbf{S}^z_n - \sum_{m=1}^L \{ \mathbf{S}^-_m , \mathbf{S}^z_m \} ,
    \label{eq:higherspinq}
\eeq
take the same form for any spin-$s$ model, with $\mathbf{S}_n^{\alpha}$ ($\alpha=\pm,z$) being the corresponding spin $s$ operator.

\section{Summary and discussions}
In this work, we found an interesting deformation in quantum many-body systems which keeps part of the spectra intact while deforming the rest. The undeformed spectra are protected by an emergent symmetry. Although easily stated, explicit constructions of such deformations have not been studied much. In the current work we rely on integrability, for obvious reasons: integrability offers us the necessary tools to find the complete spectrum. Careful analysis of the Bethe ansatz provides us clues to construct such flows. It would be very interesting to find examples with non-integrable models, but so far we still lack a clear guidance.

Even if we restrict ourselves to integrable models, there are many questions to be answered. An immediate question is how general our construction is. The spin chain models we have considered in the current work are not the most general ones. More works are needed to see whether similar constructions exist for the XXZ model when $q$ is a root of unity. There are some numerical evidences but the full structure is more subtle. Moreover, the operators $\hat{\mathfrak{q}}_\alpha$ resemble the Yangian generators in the XXX case. The algebraic origin is yet to be elucidated. Also, all the models we have considered so far preserve at least the U(1) symmetry, or magnon numbers. What about the models without U(1) symmetry such as the XYZ model~\cite{Baxter, Inami:1994wn, Fan:1996jq} or models with un-parallel boundary magnetic fields~\cite{Cao:2003sbc, Nepomechie:2003ez, Nepomechie:2003vv, Cao:2013qxa, Cao:2013bca, Zhang:2014ria, Wang_2015}? 
Those questions remain elusive, and they might reveal the underlying algebraic structure of the iso-BAE flow.

Going beyond spin chain models, it is natural to ask about other types of models such as integrable field theories with boundaries. One possible way is considering the proper continuum limit of the spin chain models~\cite{Cardy_1986, Affleck_1986, Oshikawa:1996dj, Affleck:1998nq}. For example, in the case of the ${\rm XXX}_{1/2}$ model, the field theory limit yields the $c=1$ free compactified boson BCFT at self-dual radius $R = \sqrt{2}$~\cite{Alcaraz1987, Alcaraz:1987zr}. The spectra of Neumann-Neumann and Dirichlet-Dirichlet boundary conditions coincide~\cite{Roy:2020frd}, which seems to match our expectation from iso-BAE flow. Following this route, it is highly likely that similar deformations exist for rather general integrable models.

We would like to emphasize that although the spectrum of the type-I states is invariant under the flow, the eigenvectors are deformed. This implies that even for type-I states, other physical quantities such as correlation functions~\cite{Kitanine:2007bi, Kitanine:2008wb} or entanglement measures~\cite{Laflorencie_2006, Fagotti:2010cc, Roy_2022} might vary with $\alpha$. Since the models along the iso-BAE flow are closely related to each other, this provides a new set-up to study quench dynamics~\cite{Cardy_2006, Caux_2013, Caux:2016esd} along the iso-BAE flow. Such process may be under better analytic control and shed new light on the out-of-equilibrium dynamics of integrable spin chains.

\begin{acknowledgments}
The work of Y.J. is partly supported by Startup Funding No. 3207022217A1 of Southeast University and by the NSF of China through Grant No. 12247103.
Y.M.'s work is supported by World Premier International Research Center Initiative (WPI), MEXT, Japan. We thank Jue Hou for previous collaborations on related topics. We are grateful to Philip Boyle Smith, Jie Gu, Hosho Katsura, Yifan Liu, Rafael Nepomechie, Bart Vlaar, Masahito Yamazaki and Xin Zhang for useful discussions. Y.M. would like to thank Lorentz Center for hospitality. 
\end{acknowledgments}

\bibliography{refs.bib}

\begin{thebibliography}{52}%
\makeatletter
\providecommand \@ifxundefined [1]{%
 \@ifx{#1\undefined}
}%
\providecommand \@ifnum [1]{%
 \ifnum #1\expandafter \@firstoftwo
 \else \expandafter \@secondoftwo
 \fi
}%
\providecommand \@ifx [1]{%
 \ifx #1\expandafter \@firstoftwo
 \else \expandafter \@secondoftwo
 \fi
}%
\providecommand \natexlab [1]{#1}%
\providecommand \enquote  [1]{``#1''}%
\providecommand \bibnamefont  [1]{#1}%
\providecommand \bibfnamefont [1]{#1}%
\providecommand \citenamefont [1]{#1}%
\providecommand \href@noop [0]{\@secondoftwo}%
\providecommand \href [0]{\begingroup \@sanitize@url \@href}%
\providecommand \@href[1]{\@@startlink{#1}\@@href}%
\providecommand \@@href[1]{\endgroup#1\@@endlink}%
\providecommand \@sanitize@url [0]{\catcode `\\12\catcode `\$12\catcode
  `\&12\catcode `\#12\catcode `\^12\catcode `\_12\catcode `\%12\relax}%
\providecommand \@@startlink[1]{}%
\providecommand \@@endlink[0]{}%
\providecommand \url  [0]{\begingroup\@sanitize@url \@url }%
\providecommand \@url [1]{\endgroup\@href {#1}{\urlprefix }}%
\providecommand \urlprefix  [0]{URL }%
\providecommand \Eprint [0]{\href }%
\providecommand \doibase [0]{https://doi.org/}%
\providecommand \selectlanguage [0]{\@gobble}%
\providecommand \bibinfo  [0]{\@secondoftwo}%
\providecommand \bibfield  [0]{\@secondoftwo}%
\providecommand \translation [1]{[#1]}%
\providecommand \BibitemOpen [0]{}%
\providecommand \bibitemStop [0]{}%
\providecommand \bibitemNoStop [0]{.\EOS\space}%
\providecommand \EOS [0]{\spacefactor3000\relax}%
\providecommand \BibitemShut  [1]{\csname bibitem#1\endcsname}%
\let\auto@bib@innerbib\@empty
\bibitem [{\citenamefont {Landau}\ and\ \citenamefont
  {Lifshits}(1991)}]{Landau:1991wop}%
  \BibitemOpen
  \bibfield  {author} {\bibinfo {author} {\bibfnamefont {L.~D.}\ \bibnamefont
  {Landau}}\ and\ \bibinfo {author} {\bibfnamefont {E.~M.}\ \bibnamefont
  {Lifshits}},\ }\href@noop {} {\emph {\bibinfo {title} {{Quantum Mechanics}:
  {Non-Relativistic Theory}}}},\ \bibinfo {series} {Course of Theoretical
  Physics}, Vol.~\bibinfo {volume} {3}\ (\bibinfo  {publisher}
  {Butterworth-Heinemann},\ \bibinfo {address} {Oxford},\ \bibinfo {year}
  {1991})\BibitemShut {NoStop}%
\bibitem [{\citenamefont {Cohen-Tannoudji}\ \emph {et~al.}(2019)\citenamefont
  {Cohen-Tannoudji}, \citenamefont {Diu},\ and\ \citenamefont
  {Laloë}}]{Cohen-Tannoudji:101367}%
  \BibitemOpen
  \bibfield  {author} {\bibinfo {author} {\bibfnamefont {C.}~\bibnamefont
  {Cohen-Tannoudji}}, \bibinfo {author} {\bibfnamefont {B.}~\bibnamefont
  {Diu}},\ and\ \bibinfo {author} {\bibfnamefont {F.}~\bibnamefont {Laloë}},\
  }\href {https://cds.cern.ch/record/101367} {\emph {\bibinfo {title} {{Quantum
  mechanics Volume 1}}}}\ (\bibinfo  {publisher} {Wiley},\ \bibinfo {address}
  {New York, NY},\ \bibinfo {year} {2019})\BibitemShut {NoStop}%
\bibitem [{\citenamefont {Yurov}\ and\ \citenamefont
  {Zamolodchikov}(1990)}]{Yurov:1989yu}%
  \BibitemOpen
  \bibfield  {author} {\bibinfo {author} {\bibfnamefont {V.~P.}\ \bibnamefont
  {Yurov}}\ and\ \bibinfo {author} {\bibfnamefont {A.~B.}\ \bibnamefont
  {Zamolodchikov}},\ }\bibfield  {title} {\bibinfo {title} {{Truncated
  Conformal Space Approach to Scaling Lee-Yang Model}},\ }\href
  {https://doi.org/10.1142/S0217751X9000218X} {\bibfield  {journal} {\bibinfo
  {journal} {Int. J. Mod. Phys. A}\ }\textbf {\bibinfo {volume} {5}},\ \bibinfo
  {pages} {3221} (\bibinfo {year} {1990})}\BibitemShut {NoStop}%
\bibitem [{\citenamefont {Yurov}\ and\ \citenamefont
  {Zamolodchikov}(1991)}]{Yurov:1991my}%
  \BibitemOpen
  \bibfield  {author} {\bibinfo {author} {\bibfnamefont {V.~P.}\ \bibnamefont
  {Yurov}}\ and\ \bibinfo {author} {\bibfnamefont {A.~B.}\ \bibnamefont
  {Zamolodchikov}},\ }\bibfield  {title} {\bibinfo {title} {{Truncated
  fermionic space approach to the critical 2-D Ising model with magnetic
  field}},\ }\href {https://doi.org/10.1142/S0217751X91002161} {\bibfield
  {journal} {\bibinfo  {journal} {Int. J. Mod. Phys. A}\ }\textbf {\bibinfo
  {volume} {6}},\ \bibinfo {pages} {4557} (\bibinfo {year} {1991})}\BibitemShut
  {NoStop}%
\bibitem [{\citenamefont {James}\ \emph {et~al.}(2018)\citenamefont {James},
  \citenamefont {Konik}, \citenamefont {Lecheminant}, \citenamefont
  {Robinson},\ and\ \citenamefont {Tsvelik}}]{James:2017cpc}%
  \BibitemOpen
  \bibfield  {author} {\bibinfo {author} {\bibfnamefont {A.~J.~A.}\
  \bibnamefont {James}}, \bibinfo {author} {\bibfnamefont {R.~M.}\ \bibnamefont
  {Konik}}, \bibinfo {author} {\bibfnamefont {P.}~\bibnamefont {Lecheminant}},
  \bibinfo {author} {\bibfnamefont {N.~J.}\ \bibnamefont {Robinson}},\ and\
  \bibinfo {author} {\bibfnamefont {A.~M.}\ \bibnamefont {Tsvelik}},\
  }\bibfield  {title} {\bibinfo {title} {{Non-perturbative methodologies for
  low-dimensional strongly-correlated systems: From non-abelian bosonization to
  truncated spectrum methods}},\ }\href
  {https://doi.org/10.1088/1361-6633/aa91ea} {\bibfield  {journal} {\bibinfo
  {journal} {Rept. Prog. Phys.}\ }\textbf {\bibinfo {volume} {81}},\ \bibinfo
  {pages} {046002} (\bibinfo {year} {2018})},\ \Eprint
  {https://arxiv.org/abs/1703.08421} {arXiv:1703.08421 [cond-mat.str-el]}
  \BibitemShut {NoStop}%
\bibitem [{\citenamefont {Bethe}(1931)}]{Bethe_1931}%
  \BibitemOpen
  \bibfield  {author} {\bibinfo {author} {\bibfnamefont {H.~A.}\ \bibnamefont
  {Bethe}},\ }\bibfield  {title} {\bibinfo {title} {Zur {T}heorie der
  {M}etalle. i. {E}igenwerte und {E}igenfunktionen der linearen {A}tomkette},\
  }\href {https://doi.org/10.1007/BF01341708} {\bibfield  {journal} {\bibinfo
  {journal} {Zeit. f\"ur Physik}\ }\textbf {\bibinfo {volume} {71}},\ \bibinfo
  {pages} {205} (\bibinfo {year} {1931})}\BibitemShut {NoStop}%
\bibitem [{\citenamefont {Gaudin}(2014)}]{Gaudin_2014}%
  \BibitemOpen
  \bibfield  {author} {\bibinfo {author} {\bibfnamefont {M.}~\bibnamefont
  {Gaudin}},\ }\href@noop {} {\emph {\bibinfo {title} {{The Bethe
  Wavefunction}}}}\ (\bibinfo  {publisher} {Cambridge University Press},\
  \bibinfo {year} {2014})\BibitemShut {NoStop}%
\bibitem [{\citenamefont {Korepin}\ \emph {et~al.}(1993)\citenamefont
  {Korepin}, \citenamefont {Bogoliubov},\ and\ \citenamefont
  {Izergin}}]{Korepin_1993}%
  \BibitemOpen
  \bibfield  {author} {\bibinfo {author} {\bibfnamefont {V.~E.}\ \bibnamefont
  {Korepin}}, \bibinfo {author} {\bibfnamefont {N.~M.}\ \bibnamefont
  {Bogoliubov}},\ and\ \bibinfo {author} {\bibfnamefont {A.~G.}\ \bibnamefont
  {Izergin}},\ }\href {https://doi.org/10.1017/CBO9780511628832} {\emph
  {\bibinfo {title} {Quantum Inverse Scattering Method and Correlation
  Functions}}},\ Cambridge Monographs on Mathematical Physics\ (\bibinfo
  {publisher} {Cambridge University Press},\ \bibinfo {year}
  {1993})\BibitemShut {NoStop}%
\bibitem [{\citenamefont {Nepomechie}(2020)}]{Nepomechie:2019gqt}%
  \BibitemOpen
  \bibfield  {author} {\bibinfo {author} {\bibfnamefont {R.~I.}\ \bibnamefont
  {Nepomechie}},\ }\bibfield  {title} {\bibinfo {title} {{Q-systems with
  boundary parameters}},\ }\href {https://doi.org/10.1088/1751-8121/ab9386}
  {\bibfield  {journal} {\bibinfo  {journal} {J. Phys. A}\ }\textbf {\bibinfo
  {volume} {53}},\ \bibinfo {pages} {294001} (\bibinfo {year} {2020})},\
  \Eprint {https://arxiv.org/abs/1912.12702} {arXiv:1912.12702 [hep-th]}
  \BibitemShut {NoStop}%
\bibitem [{\citenamefont {Marboe}\ and\ \citenamefont
  {Volin}(2017)}]{Marboe:2016yyn}%
  \BibitemOpen
  \bibfield  {author} {\bibinfo {author} {\bibfnamefont {C.}~\bibnamefont
  {Marboe}}\ and\ \bibinfo {author} {\bibfnamefont {D.}~\bibnamefont {Volin}},\
  }\bibfield  {title} {\bibinfo {title} {{Fast analytic solver of rational
  Bethe equations}},\ }\href {https://doi.org/10.1088/1751-8121/aa6b88}
  {\bibfield  {journal} {\bibinfo  {journal} {J. Phys. A}\ }\textbf {\bibinfo
  {volume} {50}},\ \bibinfo {pages} {204002} (\bibinfo {year} {2017})},\
  \Eprint {https://arxiv.org/abs/1608.06504} {arXiv:1608.06504 [math-ph]}
  \BibitemShut {NoStop}%
\bibitem [{\citenamefont {Granet}\ and\ \citenamefont
  {Jacobsen}(2020)}]{Granet:2019knz}%
  \BibitemOpen
  \bibfield  {author} {\bibinfo {author} {\bibfnamefont {E.}~\bibnamefont
  {Granet}}\ and\ \bibinfo {author} {\bibfnamefont {J.~L.}\ \bibnamefont
  {Jacobsen}},\ }\bibfield  {title} {\bibinfo {title} {{On zero-remainder
  conditions in the Bethe ansatz}},\ }\href
  {https://doi.org/10.1007/JHEP03(2020)178} {\bibfield  {journal} {\bibinfo
  {journal} {JHEP}\ }\textbf {\bibinfo {volume} {03}},\ \bibinfo {pages}
  {178}},\ \Eprint {https://arxiv.org/abs/1910.07797} {arXiv:1910.07797
  [hep-th]} \BibitemShut {NoStop}%
\bibitem [{\citenamefont {Bajnok}\ \emph {et~al.}(2020)\citenamefont {Bajnok},
  \citenamefont {Granet}, \citenamefont {Jacobsen},\ and\ \citenamefont
  {Nepomechie}}]{Bajnok:2019zub}%
  \BibitemOpen
  \bibfield  {author} {\bibinfo {author} {\bibfnamefont {Z.}~\bibnamefont
  {Bajnok}}, \bibinfo {author} {\bibfnamefont {E.}~\bibnamefont {Granet}},
  \bibinfo {author} {\bibfnamefont {J.~L.}\ \bibnamefont {Jacobsen}},\ and\
  \bibinfo {author} {\bibfnamefont {R.~I.}\ \bibnamefont {Nepomechie}},\
  }\bibfield  {title} {\bibinfo {title} {{On Generalized $Q$-systems}},\ }\href
  {https://doi.org/10.1007/JHEP03(2020)177} {\bibfield  {journal} {\bibinfo
  {journal} {JHEP}\ }\textbf {\bibinfo {volume} {03}},\ \bibinfo {pages}
  {177}},\ \Eprint {https://arxiv.org/abs/1910.07805} {arXiv:1910.07805
  [hep-th]} \BibitemShut {NoStop}%
\bibitem [{\citenamefont {Hou}\ \emph {et~al.}(2024{\natexlab{a}})\citenamefont
  {Hou}, \citenamefont {Jiang},\ and\ \citenamefont {Miao}}]{Hou:2023jkr}%
  \BibitemOpen
  \bibfield  {author} {\bibinfo {author} {\bibfnamefont {J.}~\bibnamefont
  {Hou}}, \bibinfo {author} {\bibfnamefont {Y.}~\bibnamefont {Jiang}},\ and\
  \bibinfo {author} {\bibfnamefont {Y.}~\bibnamefont {Miao}},\ }\bibfield
  {title} {\bibinfo {title} {{Rational Q-systems at Root of Unity I. Closed
  Chains}},\ }\href {https://doi.org/10.21468/SciPostPhys.16.5.129} {\bibfield
  {journal} {\bibinfo  {journal} {SciPost Phys.}\ }\textbf {\bibinfo {volume}
  {16}},\ \bibinfo {pages} {129} (\bibinfo {year} {2024}{\natexlab{a}})},\
  \Eprint {https://arxiv.org/abs/2310.14966} {arXiv:2310.14966 [hep-th]}
  \BibitemShut {NoStop}%
\bibitem [{\citenamefont {Hou}\ \emph {et~al.}(2024{\natexlab{b}})\citenamefont
  {Hou}, \citenamefont {Jiang},\ and\ \citenamefont {Zhu}}]{Hou:2023ndn}%
  \BibitemOpen
  \bibfield  {author} {\bibinfo {author} {\bibfnamefont {J.}~\bibnamefont
  {Hou}}, \bibinfo {author} {\bibfnamefont {Y.}~\bibnamefont {Jiang}},\ and\
  \bibinfo {author} {\bibfnamefont {R.-D.}\ \bibnamefont {Zhu}},\ }\bibfield
  {title} {\bibinfo {title} {{Spin-$s$ rational $Q$-system}},\ }\href
  {https://doi.org/10.21468/SciPostPhys.16.4.113} {\bibfield  {journal}
  {\bibinfo  {journal} {SciPost Phys.}\ }\textbf {\bibinfo {volume} {16}},\
  \bibinfo {pages} {113} (\bibinfo {year} {2024}{\natexlab{b}})},\ \Eprint
  {https://arxiv.org/abs/2303.07640} {arXiv:2303.07640 [hep-th]} \BibitemShut
  {NoStop}%
\bibitem [{\citenamefont {Alcaraz}\ \emph {et~al.}(1987)\citenamefont
  {Alcaraz}, \citenamefont {Barber}, \citenamefont {Batchelor}, \citenamefont
  {Baxter},\ and\ \citenamefont {Quispel}}]{Alcaraz1987}%
  \BibitemOpen
  \bibfield  {author} {\bibinfo {author} {\bibfnamefont {F.~C.}\ \bibnamefont
  {Alcaraz}}, \bibinfo {author} {\bibfnamefont {M.~N.}\ \bibnamefont {Barber}},
  \bibinfo {author} {\bibfnamefont {M.~T.}\ \bibnamefont {Batchelor}}, \bibinfo
  {author} {\bibfnamefont {R.~J.}\ \bibnamefont {Baxter}},\ and\ \bibinfo
  {author} {\bibfnamefont {G.~R.~W.}\ \bibnamefont {Quispel}},\ }\bibfield
  {title} {\bibinfo {title} {{Surface Exponents of the Quantum XXZ,
  Ashkin-Teller and Potts Models}},\ }\href
  {https://doi.org/10.1088/0305-4470/20/18/038} {\bibfield  {journal} {\bibinfo
   {journal} {J. Phys. A}\ }\textbf {\bibinfo {volume} {20}},\ \bibinfo {pages}
  {6397} (\bibinfo {year} {1987})}\BibitemShut {NoStop}%
\bibitem [{\citenamefont {Sklyanin}(1988)}]{Sklyanin:1988yz}%
  \BibitemOpen
  \bibfield  {author} {\bibinfo {author} {\bibfnamefont {E.~K.}\ \bibnamefont
  {Sklyanin}},\ }\bibfield  {title} {\bibinfo {title} {{Boundary Conditions for
  Integrable Quantum Systems}},\ }\href
  {https://doi.org/10.1088/0305-4470/21/10/015} {\bibfield  {journal} {\bibinfo
   {journal} {J. Phys. A}\ }\textbf {\bibinfo {volume} {21}},\ \bibinfo {pages}
  {2375} (\bibinfo {year} {1988})}\BibitemShut {NoStop}%
\bibitem [{\citenamefont {Morin-Duchesne}\ \emph
  {et~al.}(2016{\natexlab{a}})\citenamefont {Morin-Duchesne}, \citenamefont
  {Rasmussen}, \citenamefont {Ruelle},\ and\ \citenamefont
  {Saint-Aubin}}]{Aubin:2015afa}%
  \BibitemOpen
  \bibfield  {author} {\bibinfo {author} {\bibfnamefont {A.}~\bibnamefont
  {Morin-Duchesne}}, \bibinfo {author} {\bibfnamefont {J.}~\bibnamefont
  {Rasmussen}}, \bibinfo {author} {\bibfnamefont {P.}~\bibnamefont {Ruelle}},\
  and\ \bibinfo {author} {\bibfnamefont {Y.}~\bibnamefont {Saint-Aubin}},\
  }\bibfield  {title} {\bibinfo {title} {{On the reality of spectra of
  $\boldsymbol{U_q(sl_2)}$-invariant XXZ Hamiltonians}},\ }\href
  {https://doi.org/10.1088/1742-5468/2016/05/053105} {\bibfield  {journal}
  {\bibinfo  {journal} {J. Stat. Mech.}\ }\textbf {\bibinfo {volume} {1605}},\
  \bibinfo {pages} {053105} (\bibinfo {year} {2016}{\natexlab{a}})},\ \Eprint
  {https://arxiv.org/abs/1502.01859} {arXiv:1502.01859 [math-ph]} \BibitemShut
  {NoStop}%
\bibitem [{\citenamefont {Pasquier}\ and\ \citenamefont
  {Saleur}(1990)}]{Pasquier:1989kd}%
  \BibitemOpen
  \bibfield  {author} {\bibinfo {author} {\bibfnamefont {V.}~\bibnamefont
  {Pasquier}}\ and\ \bibinfo {author} {\bibfnamefont {H.}~\bibnamefont
  {Saleur}},\ }\bibfield  {title} {\bibinfo {title} {{Common Structures Between
  Finite Systems and Conformal Field Theories Through Quantum Groups}},\ }\href
  {https://doi.org/10.1016/0550-3213(90)90122-T} {\bibfield  {journal}
  {\bibinfo  {journal} {Nucl. Phys. B}\ }\textbf {\bibinfo {volume} {330}},\
  \bibinfo {pages} {523} (\bibinfo {year} {1990})}\BibitemShut {NoStop}%
\bibitem [{\citenamefont {Miao}\ \emph {et~al.}(2021)\citenamefont {Miao},
  \citenamefont {Lamers},\ and\ \citenamefont {Pasquier}}]{Miao:2020irt}%
  \BibitemOpen
  \bibfield  {author} {\bibinfo {author} {\bibfnamefont {Y.}~\bibnamefont
  {Miao}}, \bibinfo {author} {\bibfnamefont {J.}~\bibnamefont {Lamers}},\ and\
  \bibinfo {author} {\bibfnamefont {V.}~\bibnamefont {Pasquier}},\ }\bibfield
  {title} {\bibinfo {title} {{On the Q operator and the spectrum of the XXZ
  model at root of unity}},\ }\href
  {https://doi.org/10.21468/SciPostPhys.11.3.067} {\bibfield  {journal}
  {\bibinfo  {journal} {SciPost Phys.}\ }\textbf {\bibinfo {volume} {11}},\
  \bibinfo {pages} {067} (\bibinfo {year} {2021})},\ \Eprint
  {https://arxiv.org/abs/2012.10224} {arXiv:2012.10224 [cond-mat.stat-mech]}
  \BibitemShut {NoStop}%
\bibitem [{\citenamefont {Gainutdinov}\ and\ \citenamefont
  {Nepomechie}(2016)}]{Azat:2016pxy}%
  \BibitemOpen
  \bibfield  {author} {\bibinfo {author} {\bibfnamefont {A.~M.}\ \bibnamefont
  {Gainutdinov}}\ and\ \bibinfo {author} {\bibfnamefont {R.~I.}\ \bibnamefont
  {Nepomechie}},\ }\bibfield  {title} {\bibinfo {title} {{Algebraic Bethe
  ansatz for the quantum group invariant open XXZ chain at roots of unity}},\
  }\href {https://doi.org/10.1016/j.nuclphysb.2016.06.007} {\bibfield
  {journal} {\bibinfo  {journal} {Nucl. Phys. B}\ }\textbf {\bibinfo {volume}
  {909}},\ \bibinfo {pages} {796} (\bibinfo {year} {2016})},\ \Eprint
  {https://arxiv.org/abs/1603.09249} {arXiv:1603.09249 [math-ph]} \BibitemShut
  {NoStop}%
\bibitem [{\citenamefont {Takhtajan}(1982)}]{Takhtajan:1982jeo}%
  \BibitemOpen
  \bibfield  {author} {\bibinfo {author} {\bibfnamefont {L.~A.}\ \bibnamefont
  {Takhtajan}},\ }\bibfield  {title} {\bibinfo {title} {{The picture of
  low-lying excitations in the isotropic Heisenberg chain of arbitrary
  spins}},\ }\href {https://doi.org/10.1016/0375-9601(82)90764-2} {\bibfield
  {journal} {\bibinfo  {journal} {Phys. Lett. A}\ }\textbf {\bibinfo {volume}
  {87}},\ \bibinfo {pages} {479} (\bibinfo {year} {1982})}\BibitemShut
  {NoStop}%
\bibitem [{\citenamefont {Babujian}(1983)}]{Babujian:1983ae}%
  \BibitemOpen
  \bibfield  {author} {\bibinfo {author} {\bibfnamefont {H.~M.}\ \bibnamefont
  {Babujian}},\ }\bibfield  {title} {\bibinfo {title} {{Exact Solution of the
  Isotropic Heisenberg Chain with Arbitrary Spins: Thermodynamics of the
  Model}},\ }\href {https://doi.org/10.1016/0550-3213(83)90668-5} {\bibfield
  {journal} {\bibinfo  {journal} {Nucl. Phys. B}\ }\textbf {\bibinfo {volume}
  {215}},\ \bibinfo {pages} {317} (\bibinfo {year} {1983})}\BibitemShut
  {NoStop}%
\bibitem [{\citenamefont {Baxter}(1982)}]{Baxter}%
  \BibitemOpen
  \bibfield  {author} {\bibinfo {author} {\bibfnamefont {R.~J.}\ \bibnamefont
  {Baxter}},\ }\href@noop {} {\emph {\bibinfo {title} {{Exactly Solved Models
  in Statistical Mechanics}}}}\ (\bibinfo  {publisher} {Academic Press},\
  \bibinfo {year} {1982})\BibitemShut {NoStop}%
\bibitem [{\citenamefont {Inami}\ and\ \citenamefont
  {Konno}(1994)}]{Inami:1994wn}%
  \BibitemOpen
  \bibfield  {author} {\bibinfo {author} {\bibfnamefont {T.}~\bibnamefont
  {Inami}}\ and\ \bibinfo {author} {\bibfnamefont {H.}~\bibnamefont {Konno}},\
  }\bibfield  {title} {\bibinfo {title} {{Integrable XYZ spin chain with
  boundaries}},\ }\href {https://doi.org/10.1088/0305-4470/27/24/002}
  {\bibfield  {journal} {\bibinfo  {journal} {J. Phys. A}\ }\textbf {\bibinfo
  {volume} {27}},\ \bibinfo {pages} {L913} (\bibinfo {year} {1994})},\ \Eprint
  {https://arxiv.org/abs/hep-th/9409138} {arXiv:hep-th/9409138} \BibitemShut
  {NoStop}%
\bibitem [{\citenamefont {Fan}\ \emph {et~al.}(1996)\citenamefont {Fan},
  \citenamefont {Hou}, \citenamefont {Shi},\ and\ \citenamefont
  {Yang}}]{Fan:1996jq}%
  \BibitemOpen
  \bibfield  {author} {\bibinfo {author} {\bibfnamefont {H.}~\bibnamefont
  {Fan}}, \bibinfo {author} {\bibfnamefont {B.-Y.}\ \bibnamefont {Hou}},
  \bibinfo {author} {\bibfnamefont {K.-J.}\ \bibnamefont {Shi}},\ and\ \bibinfo
  {author} {\bibfnamefont {Z.-X.}\ \bibnamefont {Yang}},\ }\bibfield  {title}
  {\bibinfo {title} {{Algebraic Bethe ansatz for eight vertex model with
  general open boundary conditions}},\ }\href
  {https://doi.org/10.1016/0550-3213(96)00398-7} {\bibfield  {journal}
  {\bibinfo  {journal} {Nucl. Phys. B}\ }\textbf {\bibinfo {volume} {478}},\
  \bibinfo {pages} {723} (\bibinfo {year} {1996})},\ \Eprint
  {https://arxiv.org/abs/hep-th/9604016} {arXiv:hep-th/9604016} \BibitemShut
  {NoStop}%
\bibitem [{\citenamefont {Cao}\ \emph {et~al.}(2003)\citenamefont {Cao},
  \citenamefont {Lin}, \citenamefont {Shi},\ and\ \citenamefont
  {Wang}}]{Cao:2003sbc}%
  \BibitemOpen
  \bibfield  {author} {\bibinfo {author} {\bibfnamefont {J.}~\bibnamefont
  {Cao}}, \bibinfo {author} {\bibfnamefont {H.-Q.}\ \bibnamefont {Lin}},
  \bibinfo {author} {\bibfnamefont {K.-J.}\ \bibnamefont {Shi}},\ and\ \bibinfo
  {author} {\bibfnamefont {Y.}~\bibnamefont {Wang}},\ }\bibfield  {title}
  {\bibinfo {title} {{Exact solution of XXZ spin chain with unparallel boundary
  fields}},\ }\href {https://doi.org/10.1016/S0550-3213(03)00372-9} {\bibfield
  {journal} {\bibinfo  {journal} {Nucl. Phys. B}\ }\textbf {\bibinfo {volume}
  {663}},\ \bibinfo {pages} {487} (\bibinfo {year} {2003})}\BibitemShut
  {NoStop}%
\bibitem [{\citenamefont {Nepomechie}\ and\ \citenamefont
  {Ravanini}(2003)}]{Nepomechie:2003ez}%
  \BibitemOpen
  \bibfield  {author} {\bibinfo {author} {\bibfnamefont {R.~I.}\ \bibnamefont
  {Nepomechie}}\ and\ \bibinfo {author} {\bibfnamefont {F.}~\bibnamefont
  {Ravanini}},\ }\bibfield  {title} {\bibinfo {title} {{Completeness of the
  Bethe ansatz solution of the open XXZ chain with nondiagonal boundary
  terms}},\ }\href {https://doi.org/10.1088/0305-4470/36/45/003} {\bibfield
  {journal} {\bibinfo  {journal} {J. Phys. A}\ }\textbf {\bibinfo {volume}
  {36}},\ \bibinfo {pages} {11391} (\bibinfo {year} {2003})},\ \Eprint
  {https://arxiv.org/abs/hep-th/0307095} {arXiv:hep-th/0307095} \BibitemShut
  {NoStop}%
\bibitem [{\citenamefont {Nepomechie}(2004)}]{Nepomechie:2003vv}%
  \BibitemOpen
  \bibfield  {author} {\bibinfo {author} {\bibfnamefont {R.~I.}\ \bibnamefont
  {Nepomechie}},\ }\bibfield  {title} {\bibinfo {title} {{Bethe ansatz solution
  of the open XXZ chain with nondiagonal boundary terms}},\ }\href
  {https://doi.org/10.1088/0305-4470/37/2/012} {\bibfield  {journal} {\bibinfo
  {journal} {J. Phys. A}\ }\textbf {\bibinfo {volume} {37}},\ \bibinfo {pages}
  {433} (\bibinfo {year} {2004})},\ \Eprint
  {https://arxiv.org/abs/hep-th/0304092} {arXiv:hep-th/0304092} \BibitemShut
  {NoStop}%
\bibitem [{\citenamefont {Cao}\ \emph {et~al.}(2013{\natexlab{a}})\citenamefont
  {Cao}, \citenamefont {Yang}, \citenamefont {Shi},\ and\ \citenamefont
  {Wang}}]{Cao:2013qxa}%
  \BibitemOpen
  \bibfield  {author} {\bibinfo {author} {\bibfnamefont {J.}~\bibnamefont
  {Cao}}, \bibinfo {author} {\bibfnamefont {W.-L.}\ \bibnamefont {Yang}},
  \bibinfo {author} {\bibfnamefont {K.}~\bibnamefont {Shi}},\ and\ \bibinfo
  {author} {\bibfnamefont {Y.}~\bibnamefont {Wang}},\ }\bibfield  {title}
  {\bibinfo {title} {{Off-diagonal Bethe ansatz solution of the XXX spin-chain
  with arbitrary boundary conditions}},\ }\href
  {https://doi.org/10.1016/j.nuclphysb.2013.06.022} {\bibfield  {journal}
  {\bibinfo  {journal} {Nucl. Phys. B}\ }\textbf {\bibinfo {volume} {875}},\
  \bibinfo {pages} {152} (\bibinfo {year} {2013}{\natexlab{a}})},\ \Eprint
  {https://arxiv.org/abs/1306.1742} {arXiv:1306.1742 [math-ph]} \BibitemShut
  {NoStop}%
\bibitem [{\citenamefont {Cao}\ \emph {et~al.}(2013{\natexlab{b}})\citenamefont
  {Cao}, \citenamefont {Yang}, \citenamefont {Shi},\ and\ \citenamefont
  {Wang}}]{Cao:2013bca}%
  \BibitemOpen
  \bibfield  {author} {\bibinfo {author} {\bibfnamefont {J.}~\bibnamefont
  {Cao}}, \bibinfo {author} {\bibfnamefont {W.-L.}\ \bibnamefont {Yang}},
  \bibinfo {author} {\bibfnamefont {K.}~\bibnamefont {Shi}},\ and\ \bibinfo
  {author} {\bibfnamefont {Y.}~\bibnamefont {Wang}},\ }\bibfield  {title}
  {\bibinfo {title} {{Off-diagonal Bethe ansatz solutions of the anisotropic
  spin-1/2 chains with arbitrary boundary fields}},\ }\href
  {https://doi.org/10.1016/j.nuclphysb.2013.10.001} {\bibfield  {journal}
  {\bibinfo  {journal} {Nucl. Phys. B}\ }\textbf {\bibinfo {volume} {877}},\
  \bibinfo {pages} {152} (\bibinfo {year} {2013}{\natexlab{b}})},\ \Eprint
  {https://arxiv.org/abs/1307.2023} {arXiv:1307.2023 [cond-mat.stat-mech]}
  \BibitemShut {NoStop}%
\bibitem [{\citenamefont {Zhang}\ \emph {et~al.}(2015)\citenamefont {Zhang},
  \citenamefont {Li}, \citenamefont {Cao}, \citenamefont {Yang}, \citenamefont
  {Shi},\ and\ \citenamefont {Wang}}]{Zhang:2014ria}%
  \BibitemOpen
  \bibfield  {author} {\bibinfo {author} {\bibfnamefont {X.}~\bibnamefont
  {Zhang}}, \bibinfo {author} {\bibfnamefont {Y.-Y.}\ \bibnamefont {Li}},
  \bibinfo {author} {\bibfnamefont {J.}~\bibnamefont {Cao}}, \bibinfo {author}
  {\bibfnamefont {W.-L.}\ \bibnamefont {Yang}}, \bibinfo {author}
  {\bibfnamefont {K.}~\bibnamefont {Shi}},\ and\ \bibinfo {author}
  {\bibfnamefont {Y.}~\bibnamefont {Wang}},\ }\bibfield  {title} {\bibinfo
  {title} {{Bethe states of the XXZ spin-1/2 chain with arbitrary boundary
  fields}},\ }\href {https://doi.org/10.1016/j.nuclphysb.2015.01.022}
  {\bibfield  {journal} {\bibinfo  {journal} {Nucl. Phys. B}\ }\textbf
  {\bibinfo {volume} {893}},\ \bibinfo {pages} {70} (\bibinfo {year} {2015})},\
  \Eprint {https://arxiv.org/abs/1412.6905} {arXiv:1412.6905 [math-ph]}
  \BibitemShut {NoStop}%
\bibitem [{\citenamefont {Wang}\ \emph {et~al.}(2015)\citenamefont {Wang},
  \citenamefont {Yang}, \citenamefont {Cao},\ and\ \citenamefont
  {Shi}}]{Wang_2015}%
  \BibitemOpen
  \bibfield  {author} {\bibinfo {author} {\bibfnamefont {Y.}~\bibnamefont
  {Wang}}, \bibinfo {author} {\bibfnamefont {W.-L.}\ \bibnamefont {Yang}},
  \bibinfo {author} {\bibfnamefont {J.}~\bibnamefont {Cao}},\ and\ \bibinfo
  {author} {\bibfnamefont {K.}~\bibnamefont {Shi}},\ }\href
  {https://doi.org/10.1007/978-3-662-46756-5} {\emph {\bibinfo {title}
  {{Off-Diagonal Bethe Ansatz for Exactly Solvable Models}}}}\ (\bibinfo
  {publisher} {Springer Berlin Heidelberg},\ \bibinfo {year}
  {2015})\BibitemShut {NoStop}%
\bibitem [{\citenamefont {Bl\"ote}\ \emph {et~al.}(1986)\citenamefont
  {Bl\"ote}, \citenamefont {Cardy},\ and\ \citenamefont
  {Nightingale}}]{Cardy_1986}%
  \BibitemOpen
  \bibfield  {author} {\bibinfo {author} {\bibfnamefont {H.~W.~J.}\
  \bibnamefont {Bl\"ote}}, \bibinfo {author} {\bibfnamefont {J.~L.}\
  \bibnamefont {Cardy}},\ and\ \bibinfo {author} {\bibfnamefont {M.~P.}\
  \bibnamefont {Nightingale}},\ }\bibfield  {title} {\bibinfo {title}
  {Conformal invariance, the central charge, and universal finite-size
  amplitudes at criticality},\ }\href
  {https://doi.org/10.1103/PhysRevLett.56.742} {\bibfield  {journal} {\bibinfo
  {journal} {Phys. Rev. Lett.}\ }\textbf {\bibinfo {volume} {56}},\ \bibinfo
  {pages} {742} (\bibinfo {year} {1986})}\BibitemShut {NoStop}%
\bibitem [{\citenamefont {Affleck}(1986)}]{Affleck_1986}%
  \BibitemOpen
  \bibfield  {author} {\bibinfo {author} {\bibfnamefont {I.}~\bibnamefont
  {Affleck}},\ }\bibfield  {title} {\bibinfo {title} {Universal term in the
  free energy at a critical point and the conformal anomaly},\ }\href
  {https://doi.org/10.1103/PhysRevLett.56.746} {\bibfield  {journal} {\bibinfo
  {journal} {Phys. Rev. Lett.}\ }\textbf {\bibinfo {volume} {56}},\ \bibinfo
  {pages} {746} (\bibinfo {year} {1986})}\BibitemShut {NoStop}%
\bibitem [{\citenamefont {Oshikawa}\ and\ \citenamefont
  {Affleck}(1997)}]{Oshikawa:1996dj}%
  \BibitemOpen
  \bibfield  {author} {\bibinfo {author} {\bibfnamefont {M.}~\bibnamefont
  {Oshikawa}}\ and\ \bibinfo {author} {\bibfnamefont {I.}~\bibnamefont
  {Affleck}},\ }\bibfield  {title} {\bibinfo {title} {{Boundary conformal field
  theory approach to the critical two-dimensional Ising model with a defect
  line}},\ }\href {https://doi.org/10.1016/S0550-3213(97)00219-8} {\bibfield
  {journal} {\bibinfo  {journal} {Nucl. Phys. B}\ }\textbf {\bibinfo {volume}
  {495}},\ \bibinfo {pages} {533} (\bibinfo {year} {1997})},\ \Eprint
  {https://arxiv.org/abs/cond-mat/9612187} {arXiv:cond-mat/9612187}
  \BibitemShut {NoStop}%
\bibitem [{\citenamefont {Affleck}\ \emph {et~al.}(1998)\citenamefont
  {Affleck}, \citenamefont {Oshikawa},\ and\ \citenamefont
  {Saleur}}]{Affleck:1998nq}%
  \BibitemOpen
  \bibfield  {author} {\bibinfo {author} {\bibfnamefont {I.}~\bibnamefont
  {Affleck}}, \bibinfo {author} {\bibfnamefont {M.}~\bibnamefont {Oshikawa}},\
  and\ \bibinfo {author} {\bibfnamefont {H.}~\bibnamefont {Saleur}},\
  }\bibfield  {title} {\bibinfo {title} {{Boundary critical phenomena in the
  three state Potts model}},\ }\href
  {https://doi.org/10.1088/0305-4470/31/28/003} {\bibfield  {journal} {\bibinfo
   {journal} {J. Phys. A}\ }\textbf {\bibinfo {volume} {31}},\ \bibinfo {pages}
  {5827} (\bibinfo {year} {1998})},\ \Eprint
  {https://arxiv.org/abs/cond-mat/9804117} {arXiv:cond-mat/9804117}
  \BibitemShut {NoStop}%
\bibitem [{\citenamefont {Alcaraz}\ \emph {et~al.}(1988)\citenamefont
  {Alcaraz}, \citenamefont {Barber},\ and\ \citenamefont
  {Batchelor}}]{Alcaraz:1987zr}%
  \BibitemOpen
  \bibfield  {author} {\bibinfo {author} {\bibfnamefont {F.~C.}\ \bibnamefont
  {Alcaraz}}, \bibinfo {author} {\bibfnamefont {M.~N.}\ \bibnamefont
  {Barber}},\ and\ \bibinfo {author} {\bibfnamefont {M.~T.}\ \bibnamefont
  {Batchelor}},\ }\bibfield  {title} {\bibinfo {title} {{Conformal Invariance,
  the XXZ Chain and the Operator Content of Two-dimensional Critical
  Systems}},\ }\href {https://doi.org/10.1016/0003-4916(88)90015-2} {\bibfield
  {journal} {\bibinfo  {journal} {Annals Phys.}\ }\textbf {\bibinfo {volume}
  {182}},\ \bibinfo {pages} {280} (\bibinfo {year} {1988})}\BibitemShut
  {NoStop}%
\bibitem [{\citenamefont {Roy}\ \emph {et~al.}(2020)\citenamefont {Roy},
  \citenamefont {Pollmann},\ and\ \citenamefont {Saleur}}]{Roy:2020frd}%
  \BibitemOpen
  \bibfield  {author} {\bibinfo {author} {\bibfnamefont {A.}~\bibnamefont
  {Roy}}, \bibinfo {author} {\bibfnamefont {F.}~\bibnamefont {Pollmann}},\ and\
  \bibinfo {author} {\bibfnamefont {H.}~\bibnamefont {Saleur}},\ }\bibfield
  {title} {\bibinfo {title} {{Entanglement Hamiltonian of the 1+1-dimensional
  free, compactified boson conformal field theory}},\ }\href
  {https://doi.org/10.1088/1742-5468/aba498} {\bibfield  {journal} {\bibinfo
  {journal} {J. Stat. Mech.}\ }\textbf {\bibinfo {volume} {2008}},\ \bibinfo
  {pages} {083104} (\bibinfo {year} {2020})},\ \Eprint
  {https://arxiv.org/abs/2004.14370} {arXiv:2004.14370 [cond-mat.stat-mech]}
  \BibitemShut {NoStop}%
\bibitem [{\citenamefont {Kitanine}\ \emph {et~al.}(2007)\citenamefont
  {Kitanine}, \citenamefont {Kozlowski}, \citenamefont {Maillet}, \citenamefont
  {Niccoli}, \citenamefont {Slavnov},\ and\ \citenamefont
  {Terras}}]{Kitanine:2007bi}%
  \BibitemOpen
  \bibfield  {author} {\bibinfo {author} {\bibfnamefont {N.}~\bibnamefont
  {Kitanine}}, \bibinfo {author} {\bibfnamefont {K.~K.}\ \bibnamefont
  {Kozlowski}}, \bibinfo {author} {\bibfnamefont {J.~M.}\ \bibnamefont
  {Maillet}}, \bibinfo {author} {\bibfnamefont {G.}~\bibnamefont {Niccoli}},
  \bibinfo {author} {\bibfnamefont {N.~A.}\ \bibnamefont {Slavnov}},\ and\
  \bibinfo {author} {\bibfnamefont {V.}~\bibnamefont {Terras}},\ }\bibfield
  {title} {\bibinfo {title} {{Correlation functions of the open XXZ chain I}},\
  }\href {https://doi.org/10.1088/1742-5468/2007/10/P10009} {\bibfield
  {journal} {\bibinfo  {journal} {J. Stat. Mech.}\ }\textbf {\bibinfo {volume}
  {0710}},\ \bibinfo {pages} {P10009} (\bibinfo {year} {2007})},\ \Eprint
  {https://arxiv.org/abs/0707.1995} {arXiv:0707.1995 [hep-th]} \BibitemShut
  {NoStop}%
\bibitem [{\citenamefont {Kitanine}\ \emph {et~al.}(2008)\citenamefont
  {Kitanine}, \citenamefont {Kozlowski}, \citenamefont {Maillet}, \citenamefont
  {Niccoli}, \citenamefont {Slavnov},\ and\ \citenamefont
  {Terras}}]{Kitanine:2008wb}%
  \BibitemOpen
  \bibfield  {author} {\bibinfo {author} {\bibfnamefont {N.}~\bibnamefont
  {Kitanine}}, \bibinfo {author} {\bibfnamefont {K.~K.}\ \bibnamefont
  {Kozlowski}}, \bibinfo {author} {\bibfnamefont {J.~M.}\ \bibnamefont
  {Maillet}}, \bibinfo {author} {\bibfnamefont {G.}~\bibnamefont {Niccoli}},
  \bibinfo {author} {\bibfnamefont {N.~A.}\ \bibnamefont {Slavnov}},\ and\
  \bibinfo {author} {\bibfnamefont {V.}~\bibnamefont {Terras}},\ }\bibfield
  {title} {\bibinfo {title} {{Correlation functions of the open XXZ chain
  II}},\ }\href {https://doi.org/10.1088/1742-5468/2008/07/P07010} {\bibfield
  {journal} {\bibinfo  {journal} {J. Stat. Mech.}\ }\textbf {\bibinfo {volume}
  {0807}},\ \bibinfo {pages} {P07010} (\bibinfo {year} {2008})},\ \Eprint
  {https://arxiv.org/abs/0803.3305} {arXiv:0803.3305 [hep-th]} \BibitemShut
  {NoStop}%
\bibitem [{\citenamefont {Laflorencie}\ \emph {et~al.}(2006)\citenamefont
  {Laflorencie}, \citenamefont {S\o{}rensen}, \citenamefont {Chang},\ and\
  \citenamefont {Affleck}}]{Laflorencie_2006}%
  \BibitemOpen
  \bibfield  {author} {\bibinfo {author} {\bibfnamefont {N.}~\bibnamefont
  {Laflorencie}}, \bibinfo {author} {\bibfnamefont {E.~S.}\ \bibnamefont
  {S\o{}rensen}}, \bibinfo {author} {\bibfnamefont {M.-S.}\ \bibnamefont
  {Chang}},\ and\ \bibinfo {author} {\bibfnamefont {I.}~\bibnamefont
  {Affleck}},\ }\bibfield  {title} {\bibinfo {title} {{Boundary Effects in the
  Critical Scaling of Entanglement Entropy in 1D Systems}},\ }\href
  {https://doi.org/10.1103/PhysRevLett.96.100603} {\bibfield  {journal}
  {\bibinfo  {journal} {Phys. Rev. Lett.}\ }\textbf {\bibinfo {volume} {96}},\
  \bibinfo {pages} {100603} (\bibinfo {year} {2006})}\BibitemShut {NoStop}%
\bibitem [{\citenamefont {Fagotti}\ and\ \citenamefont
  {Calabrese}(2011)}]{Fagotti:2010cc}%
  \BibitemOpen
  \bibfield  {author} {\bibinfo {author} {\bibfnamefont {M.}~\bibnamefont
  {Fagotti}}\ and\ \bibinfo {author} {\bibfnamefont {P.}~\bibnamefont
  {Calabrese}},\ }\bibfield  {title} {\bibinfo {title} {{Universal parity
  effects in the entanglement entropy of XX chains with open boundary
  conditions}},\ }\href {https://doi.org/10.1088/1742-5468/2011/01/P01017}
  {\bibfield  {journal} {\bibinfo  {journal} {J. Stat. Mech.}\ }\textbf
  {\bibinfo {volume} {1101}},\ \bibinfo {pages} {P01017} (\bibinfo {year}
  {2011})},\ \Eprint {https://arxiv.org/abs/1010.5796} {arXiv:1010.5796
  [cond-mat.stat-mech]} \BibitemShut {NoStop}%
\bibitem [{\citenamefont {Roy}\ and\ \citenamefont {Saleur}(2022)}]{Roy_2022}%
  \BibitemOpen
  \bibfield  {author} {\bibinfo {author} {\bibfnamefont {A.}~\bibnamefont
  {Roy}}\ and\ \bibinfo {author} {\bibfnamefont {H.}~\bibnamefont {Saleur}},\
  }\bibinfo {title} {{Entanglement Entropy in Critical Quantum Spin Chains with
  Boundaries and Defects}},\ in\ \href
  {https://doi.org/10.1007/978-3-031-03998-0_3} {\emph {\bibinfo {booktitle}
  {Entanglement in Spin Chains}}}\ (\bibinfo  {publisher} {Springer
  International Publishing},\ \bibinfo {year} {2022})\ p.\ \bibinfo {pages}
  {41–60}\BibitemShut {NoStop}%
\bibitem [{\citenamefont {Calabrese}\ and\ \citenamefont
  {Cardy}(2006)}]{Cardy_2006}%
  \BibitemOpen
  \bibfield  {author} {\bibinfo {author} {\bibfnamefont {P.}~\bibnamefont
  {Calabrese}}\ and\ \bibinfo {author} {\bibfnamefont {J.}~\bibnamefont
  {Cardy}},\ }\bibfield  {title} {\bibinfo {title} {{Time Dependence of
  Correlation Functions Following a Quantum Quench}},\ }\href
  {https://doi.org/10.1103/PhysRevLett.96.136801} {\bibfield  {journal}
  {\bibinfo  {journal} {Phys. Rev. Lett.}\ }\textbf {\bibinfo {volume} {96}},\
  \bibinfo {pages} {136801} (\bibinfo {year} {2006})}\BibitemShut {NoStop}%
\bibitem [{\citenamefont {Caux}\ and\ \citenamefont
  {Essler}(2013)}]{Caux_2013}%
  \BibitemOpen
  \bibfield  {author} {\bibinfo {author} {\bibfnamefont {J.-S.}\ \bibnamefont
  {Caux}}\ and\ \bibinfo {author} {\bibfnamefont {F.~H.~L.}\ \bibnamefont
  {Essler}},\ }\bibfield  {title} {\bibinfo {title} {Time evolution of local
  observables after quenching to an integrable model},\ }\href
  {https://doi.org/10.1103/PhysRevLett.110.257203} {\bibfield  {journal}
  {\bibinfo  {journal} {Phys. Rev. Lett.}\ }\textbf {\bibinfo {volume} {110}},\
  \bibinfo {pages} {257203} (\bibinfo {year} {2013})}\BibitemShut {NoStop}%
\bibitem [{\citenamefont {Caux}(2016)}]{Caux:2016esd}%
  \BibitemOpen
  \bibfield  {author} {\bibinfo {author} {\bibfnamefont {J.-S.}\ \bibnamefont
  {Caux}},\ }\bibfield  {title} {\bibinfo {title} {{The Quench Action}},\
  }\href {https://doi.org/10.1088/1742-5468/2016/06/064006} {\bibfield
  {journal} {\bibinfo  {journal} {J. Stat. Mech.}\ }\textbf {\bibinfo {volume}
  {1606}},\ \bibinfo {pages} {064006} (\bibinfo {year} {2016})},\ \Eprint
  {https://arxiv.org/abs/1603.04689} {arXiv:1603.04689 [cond-mat.str-el]}
  \BibitemShut {NoStop}%
\bibitem [{\citenamefont {Kulish}\ \emph {et~al.}(1981)\citenamefont {Kulish},
  \citenamefont {Reshetikhin},\ and\ \citenamefont {Sklyanin}}]{Kulish:1981gi}%
  \BibitemOpen
  \bibfield  {author} {\bibinfo {author} {\bibfnamefont {P.~P.}\ \bibnamefont
  {Kulish}}, \bibinfo {author} {\bibfnamefont {N.~Y.}\ \bibnamefont
  {Reshetikhin}},\ and\ \bibinfo {author} {\bibfnamefont {E.~K.}\ \bibnamefont
  {Sklyanin}},\ }\bibfield  {title} {\bibinfo {title} {{Yang-Baxter Equation
  and Representation Theory. 1.}},\ }\href {https://doi.org/10.1007/BF02285311}
  {\bibfield  {journal} {\bibinfo  {journal} {Lett. Math. Phys.}\ }\textbf
  {\bibinfo {volume} {5}},\ \bibinfo {pages} {393} (\bibinfo {year}
  {1981})}\BibitemShut {NoStop}%
\bibitem [{\citenamefont {Frassek}\ and\ \citenamefont
  {Sz\'ecs\'enyi}(2015)}]{Frassek:2015mra}%
  \BibitemOpen
  \bibfield  {author} {\bibinfo {author} {\bibfnamefont {R.}~\bibnamefont
  {Frassek}}\ and\ \bibinfo {author} {\bibfnamefont {I.~M.}\ \bibnamefont
  {Sz\'ecs\'enyi}},\ }\bibfield  {title} {\bibinfo {title} {{Q-operators for
  the open Heisenberg spin chain}},\ }\href
  {https://doi.org/10.1016/j.nuclphysb.2015.10.010} {\bibfield  {journal}
  {\bibinfo  {journal} {Nucl. Phys. B}\ }\textbf {\bibinfo {volume} {901}},\
  \bibinfo {pages} {229} (\bibinfo {year} {2015})},\ \Eprint
  {https://arxiv.org/abs/1509.04867} {arXiv:1509.04867 [math-ph]} \BibitemShut
  {NoStop}%
\bibitem [{\citenamefont {{Vlaar}}\ and\ \citenamefont
  {{Weston}}(2020)}]{Vlaar2020}%
  \BibitemOpen
  \bibfield  {author} {\bibinfo {author} {\bibfnamefont {B.}~\bibnamefont
  {{Vlaar}}}\ and\ \bibinfo {author} {\bibfnamefont {R.}~\bibnamefont
  {{Weston}}},\ }\bibfield  {title} {\bibinfo {title} {{A Q-operator for open
  spin chains I. Baxter's TQ relation}},\ }\href
  {https://doi.org/10.1088/1751-8121/ab8854} {\bibfield  {journal} {\bibinfo
  {journal} {Journal of Physics A Mathematical General}\ }\textbf {\bibinfo
  {volume} {53}},\ \bibinfo {eid} {245205} (\bibinfo {year} {2020})},\ \Eprint
  {https://arxiv.org/abs/2001.10760} {arXiv:2001.10760 [math-ph]} \BibitemShut
  {NoStop}%
\bibitem [{\citenamefont {Frassek}\ and\ \citenamefont
  {Sz\'ecs\'enyi}(2022)}]{Frassek:2022izn}%
  \BibitemOpen
  \bibfield  {author} {\bibinfo {author} {\bibfnamefont {R.}~\bibnamefont
  {Frassek}}\ and\ \bibinfo {author} {\bibfnamefont {I.~M.}\ \bibnamefont
  {Sz\'ecs\'enyi}},\ }\bibfield  {title} {\bibinfo {title} {{Algebraic Bethe
  ansatz for Q-operators of the open XXX Heisenberg chain with arbitrary
  spin}},\ }\href {https://doi.org/10.1088/1751-8121/aca5d3} {\bibfield
  {journal} {\bibinfo  {journal} {J. Phys. A}\ }\textbf {\bibinfo {volume}
  {55}},\ \bibinfo {pages} {505201} (\bibinfo {year} {2022})},\ \Eprint
  {https://arxiv.org/abs/2208.02197} {arXiv:2208.02197 [math-ph]} \BibitemShut
  {NoStop}%
\bibitem [{\citenamefont {Morin-Duchesne}\ \emph
  {et~al.}(2016{\natexlab{b}})\citenamefont {Morin-Duchesne}, \citenamefont
  {Rasmussen}, \citenamefont {Ruelle},\ and\ \citenamefont
  {Saint-Aubin}}]{Morin-Duchesne:2015afa}%
  \BibitemOpen
  \bibfield  {author} {\bibinfo {author} {\bibfnamefont {A.}~\bibnamefont
  {Morin-Duchesne}}, \bibinfo {author} {\bibfnamefont {J.}~\bibnamefont
  {Rasmussen}}, \bibinfo {author} {\bibfnamefont {P.}~\bibnamefont {Ruelle}},\
  and\ \bibinfo {author} {\bibfnamefont {Y.}~\bibnamefont {Saint-Aubin}},\
  }\bibfield  {title} {\bibinfo {title} {{On the reality of spectra of
  $\boldsymbol{U_q(sl_2)}$-invariant XXZ Hamiltonians}},\ }\href
  {https://doi.org/10.1088/1742-5468/2016/05/053105} {\bibfield  {journal}
  {\bibinfo  {journal} {J. Stat. Mech.}\ }\textbf {\bibinfo {volume} {1605}},\
  \bibinfo {pages} {053105} (\bibinfo {year} {2016}{\natexlab{b}})},\ \Eprint
  {https://arxiv.org/abs/1502.01859} {arXiv:1502.01859 [math-ph]} \BibitemShut
  {NoStop}%
\bibitem [{\citenamefont {Kattel}\ \emph {et~al.}(2024)\citenamefont {Kattel},
  \citenamefont {Pasnoori}, \citenamefont {Pixley}, \citenamefont {Azaria},\
  and\ \citenamefont {Andrei}}]{Kattel:2023mfu}%
  \BibitemOpen
  \bibfield  {author} {\bibinfo {author} {\bibfnamefont {P.}~\bibnamefont
  {Kattel}}, \bibinfo {author} {\bibfnamefont {P.~R.}\ \bibnamefont
  {Pasnoori}}, \bibinfo {author} {\bibfnamefont {J.~H.}\ \bibnamefont
  {Pixley}}, \bibinfo {author} {\bibfnamefont {P.}~\bibnamefont {Azaria}},\
  and\ \bibinfo {author} {\bibfnamefont {N.}~\bibnamefont {Andrei}},\
  }\bibfield  {title} {\bibinfo {title} {{Kondo effect in the isotropic
  Heisenberg spin chain}},\ }\href
  {https://doi.org/10.1103/PhysRevB.109.174416} {\bibfield  {journal} {\bibinfo
   {journal} {Phys. Rev. B}\ }\textbf {\bibinfo {volume} {109}},\ \bibinfo
  {pages} {174416} (\bibinfo {year} {2024})},\ \Eprint
  {https://arxiv.org/abs/2311.10569} {arXiv:2311.10569 [cond-mat.str-el]}
  \BibitemShut {NoStop}%
\end{thebibliography}%

\clearpage

\onecolumngrid

\begin{appendix}

\begin{center}
	\textbf{{\large Supplemental Material for\\ 
	``Spectrum-preserving Deformations of Integrable Spin Chains with Open Boundaries''}}\\
	Yunfeng Jiang and Yuan Miao
\end{center}

\section{Construction of the \texorpdfstring{$\mathcal{Q}_{\alpha}$}{Q\textalpha} operator}
\label{app:ABA}
In this section, we discuss the construction and prove the main properties of the operator $\mathcal{Q}_{\alpha}$ for the XXX$_{1/2}$, XXZ and XXX$_{s}$ spin chains.

\subsection{XXX\texorpdfstring{$_{1/2}$}{1/2} model}
\label{app:ABAXXX}
\paragraph{Algebraic Bethe ansatz}
Algebraic Bethe ansatz (ABA) provides us the necessary tools to study iso-BAE flows. Therefore we first review the ABA for integrable chains with open boundary conditions. For XXX$_{1/2}$ spin chain, we start with the rational $R$-matrix \cite{Baxter, Gaudin_2014}
\beq 
    \mathbf{R}_{ab} (u) = \bp u+\ri & {\color{gray} 0 } & {\color{gray} 0 } & {\color{gray} 0 } \\ {\color{gray} 0 } & u & \ri & {\color{gray} 0 } \\  {\color{gray} 0 } & \ri & u & {\color{gray} 0 } \\ {\color{gray} 0 } & {\color{gray} 0 }& {\color{gray} 0 } & u+i \ep ,
\eeq 
which satisfies the Yang-Baxter equation
\beq 
    \mathbf{R}_{ab} (u-v) \mathbf{R}_{ac} (u) \mathbf{R}_{b c} (v) = \mathbf{R}_{b c} (v) \mathbf{R}_{ac} (u) \mathbf{R}_{ab} (u-v) .
    \label{eq:YBE}
\eeq 
In order to describe open boundaries, we also need the boundary $K$-matrices \cite{Sklyanin:1988yz}
\beq 
    \mathbf{K}_{\mathrm{L} ,a} (u , \beta ) = \bp \ri (\beta - \frac{1}{2}) - u & {\color{gray} 0 } \\ {\color{gray} 0 } & \ri (\beta + \frac{1}{2}) + u \ep_a , \quad \mathbf{K}_{\mathrm{R} ,a} (u , \alpha ) = \bp \ri (\alpha - \frac{1}{2}) + u & {\color{gray} 0 } \\ {\color{gray} 0 } & \ri (\alpha + \frac{1}{2}) - u \ep_a .
    \label{eq:boundaryKXXX}
\eeq 
where L and R stand for the left end and right boundaries. Here we restrict to diagonal boundary conditions where the $K$-matrices are diagonal.
The $K$-matrices satisfy the boundary Yang-Baxter equation \cite{Sklyanin:1988yz}
\beq 
\begin{split}
    & \mathbf{R}_{ab} (u-v) \mathbf{K}_{\mathrm{L} ,a} (u , \beta ) \mathbf{R}_{ba} (u+v) \mathbf{K}_{\mathrm{L} ,v} (v , \beta ) = \mathbf{K}_{\mathrm{L} ,v} (v , \beta )\mathbf{R}_{ba} (u+v) \mathbf{K}_{\mathrm{L} ,a} (u , \beta )  \mathbf{R}_{ab} (u-v) , \\
    & \mathbf{R}_{ab} (-u+v) \mathbf{K}^{t}_{\mathrm{R} ,a} (u , \alpha ) \mathbf{R}_{ab}^{t} (-u-v-2\ri ) \mathbf{K}^{t}_{\mathrm{R} ,v} (v , \alpha ) = \mathbf{K}^{t}_{\mathrm{R} ,v} (v , \alpha )\mathbf{R}^{t}_{ba} (-u-v-2\ri ) \mathbf{K}^{t}_{\mathrm{R} ,a} (u , \alpha )  \mathbf{R}_{ab} (-u+v) . 
\end{split}
\label{eq:BYBE}
\eeq 
The monodromy matrix for the open chain is defined as
\beq 
\begin{split}
    \mathbf{M}_a (u, \alpha , \beta) & =  \mathbf{K}_{\mathrm{L} ,a} (u , \beta )  \prod_{n=1}^L \mathbf{R}_{a n} (u-\ri/2 ) \mathbf{K}_{\mathrm{R} ,a} (u , \alpha ) \prod_{n=L}^1 \mathbf{R}_{a n} (u-\ri/2 ) \\ 
    & = \mathbf{K}_{\mathrm{L} ,a} (u , \beta )  \mathscr{M}_a(u,\alpha) \\
    & = \mathbf{K}_{\mathrm{L} ,a} (u , \beta )   \bp \mathscr{A} (u , \alpha) &  \mathscr{B} (u , \alpha ) \\  \mathscr{C} (u , \alpha ) &  \mathscr{D} (u , \alpha ) \ep_a.
\end{split}
\eeq 
Taking the trace in the auxiliary space gives the double-row transfer matrix 
\beq 
    \mathbf{T}_{1/2} (u, \alpha, \beta ) = \mathrm{Tr}_a  \mathbf{M}_a (u, \alpha , \beta) = (-u+\ri (\beta - \tfrac{1}{2})) \mathscr{A} (u , \alpha ) + (u+\ri (\beta + \tfrac{1}{2})) \mathscr{D} (u , \alpha ) .
\eeq 
The subscript $1/2$ denotes that the auxiliary space $a$ is in the spin-$1/2$ representation of the SU(2) algebra.\par

From the Yang-Baxter and boundary Yang-Baxter equations \eqref{eq:YBE} and \eqref{eq:BYBE}, we have
\beq 
    \mathbf{R}_{ab} (u-v)  \mathscr{M}_a(u,\alpha) \mathbf{R}_{ab}(u+v-\ri)  \mathscr{M}_b(v,\alpha) =  \mathscr{M}_b(v,\alpha) \mathbf{R}_{ab}(u+v-\ri) \mathscr{M}_a(u,\alpha) \mathbf{R}_{ab} (u-v) ,
    \label{eq:RMRM}
\eeq 
which gives a set of algebra between the quantum operators $\mathscr{A}$, $\mathscr{B}$, $\mathscr{C}$, and $\mathscr{D}$. For example, we have
\beq 
    \mathscr{B} (u, \alpha) \mathscr{A} (v , \alpha) - \frac{(u+v)(u-v)}{(u-v+\ri) (u+v-\ri )} \mathscr{A} (v , \alpha) \mathscr{B} (u, \alpha)  =  \frac{\ri \mathscr{B} (v, \alpha)\mathscr{A} (u, \alpha) }{u-v+\ri} + \frac{\ri (u-v) \mathscr{B} (v, \alpha)\mathscr{D} (u, \alpha) }{(u-v+\ri) (u+v-\ri )}  ,
    \label{eq:BA1}
\eeq 
\beq 
\begin{split}
    \mathscr{D} (u, \alpha) \mathscr{B} (v , \alpha) - & \frac{(u-v+\ri) (u+v-\ri )}{(u+v)(u-v)} \mathscr{B} (v , \alpha) \mathscr{D} (u, \alpha)  =  \frac{\mathscr{A} (u, \alpha)\mathscr{B} (v, \alpha) }{(u+v-\ri)(u-v)}-\frac{\ri (u+v)\mathscr{B} (u, \alpha)\mathscr{D} (v, \alpha) }{(u+v-\ri)(u-v)} \\
    & + \frac{\ri (u-v+\ri)^2}{(u-v)^2 (u+v)} \mathscr{B} (u, \alpha)\mathscr{A} (v, \alpha) - \frac{\ri \mathscr{B}(v,\alpha) \mathscr{A} (u, \alpha ) }{(u-v)^2 (u+v)} .
\end{split}
    \label{eq:BD1}
\eeq 

Using \eqref{eq:RMRM}, one can prove that the transfer matrices are in involution,
\beq 
    \left[ \mathbf{T}_{1/2} (u, \alpha, \beta ) , \mathbf{T}_{1/2} (v, \alpha, \beta ) \right] = 0 ,\quad \forall u,v \in \mathbb{C}.
    \label{eq:commTT}
\eeq 
The eigenvalues of the transfer matrix is the $T$-polynomial that appears in the $TQ$-relation \eqref{eq:TQrel}. As a result of \eqref{eq:commTT}, the transfer matrix can be seen as a generating function of conserved charges of the model. In particular, the Hamiltonian \eqref{eq:Hamiltonian} can be found as the logarithmic derivative of the transfer matrix,
\beq 
    \mathbf{H}_{\alpha, \beta} = -\ri \left. \partial_u \log \mathbf{T}_{1/2} (u, \alpha, \beta ) \right|_{u = 0}.
\eeq 
More explicitly, we have
\begin{align}
\mathbf{H}_{\alpha,\beta}= \sum_{n=1}^{L-1} \left( \vec{\sigma}_n \cdot \vec{\sigma}_{n+1} - 1 \right) + \frac{\sigma^z_L - 1}{\alpha} - \frac{\sigma^z_1 - 1 }{\beta}.
\label{eq:Hamiltonian2}
\end{align}
The eigenstates of the Hamiltonian \eqref{eq:Hamiltonian2} or the transfer matrix \eqref{eq:commTT} can be constructed using the off-diagonal element of the monodromy matrix,
\beq 
    | \{ u_j \}_{j=1}^M \rangle_{\alpha} = \prod_{j=1}^M \mathscr{B} (u_j , \alpha ) | \Uparrow \rangle ,
\eeq 
where $u_j$ are the Bethe roots that solve the BAE \eqref{eq:BAE} and $TQ$ relations \eqref{eq:TQrel}. 
{The eigenvalue of the transfer matrix for a the eigenstate $| \{ u_j \}_{j=1}^M \rangle_{\alpha}$ is
\beq 
    {T}_{1/2}(u,\alpha,\beta)  = - \frac{\big( u + \frac{\ri}{2} \big)^{2L+1} g \big( u - \frac{\ri}{2} \big) Q(u-\ri ) + \big( u - \frac{\ri}{2} \big)^{2L+1} f \big( u + \frac{\ri}{2} \big) Q(u + \ri ) }{u Q(u)},
\eeq 
where $Q(u)$ is the $Q$-polynomial with zeros at $\pm u_j$ associated with the eigenstate $| \{ u_j \}_{j=1}^M \rangle_{\alpha}$.
}

\paragraph{The operator $\mathcal{Q}_{\alpha}$}
\label{app:degenerateinf}

It turns out that the operator $\hat{\mathfrak{q}}_{\alpha}$ and $\mathcal{Q}_{\alpha}$ are intimately related to the expansion of the monodromy matrix elements at $u=\infty$. More precisely, the operators $\mathscr{A}$, $\mathscr{B}$, $\mathscr{C}$, $\mathscr{D}$ have the following expansion at $u\to \infty$,
\beq 
\begin{split}
    & \mathscr{A} (u,\alpha) = u^{2L+1} \left( \mathscr{A}_{\infty}^{(1)} + \frac{1}{u} \mathscr{A}_{\infty}^{(2)} + O\big( \frac{1}{u^2} \big) \right) , \\
    & \mathscr{D} (u,\alpha) = u^{2L+1} \left( \mathscr{D}_{\infty}^{(1)} + \frac{1}{u} \mathscr{D}_{\infty}^{(2)} + O\big( \frac{1}{u^2} \big) \right) , \\
    & \mathscr{B} (u,\alpha) = u^{2L-1} \left( \mathscr{B}_{\infty}^{(1)} + O\big( \frac{1}{u} \big) \right) , \\
    & \mathscr{C} (u,\alpha) = u^{2L-1} \left( \mathscr{C}_{\infty}^{(1)} + O\big( \frac{1}{u} \big) \right) .
\end{split}
\eeq 

From the definition of the monodromy matrix, it is straightforward to obtain
\beq 
    \mathscr{A}_{\infty}^{(1)} = 1 , \quad \mathscr{A}_{\infty}^{(2)} (\alpha) = 2 \ri \mathbf{S}^z + \ri \big( \alpha - \tfrac{1}{2} \big) ,
\eeq 
\beq 
    \mathscr{D}_{\infty}^{(1)} = - 1 , \quad \mathscr{D}_{\infty}^{(2)} (\alpha) = 2 \ri \mathbf{S}^z + \ri \big( \alpha + \tfrac{1}{2} \big) ,
\eeq 
\beq 
    \mathscr{B}_{\infty}^{(1)} (\alpha) = - 2 \left( \alpha \mathbf{S}^- + \sum_{m<n} \sigma^-_m \sigma^z_n \right) ,
\eeq 
\beq 
    \mathscr{C}_{\infty}^{(1)} (\alpha) = 2 \left( \alpha \mathbf{S}^+ + \sum_{m<n} \sigma^+_m \sigma^z_n \right) .
\eeq 
When $\alpha \in \mathbb{R}$, we have $(\mathscr{C}_{\infty}^{(1)}(\alpha) )^\dag = - \mathscr{B}_{\infty}^{(1)} (\alpha)$. Importantly, the operator $\hat{\mathfrak{q}}_{\alpha}$ in \eqref{eq:spinhalfq} is identified with  $\mathscr{B}_{\infty}^{(1)} (\alpha)$, \emph{i.e.}
\beq 
    \hat{\mathfrak{q}}_\alpha = \mathscr{B}_{\infty}^{(1)} (\alpha) = - 2 \left( \alpha \mathbf{S}^- + \sum_{m<n} \sigma^-_m \sigma^z_n \right) .
\eeq 

\paragraph{Commutativity of $\mathcal{Q}_{\alpha}$ and the transfer matrix} To prove the commutativity of $\mathcal{Q}_{\alpha}$ and the transfer matrix, we need the commutation relation of $\hat{\mathfrak{q}}_{\alpha}$ and the monodromy matrix elements. This can be achieved by taking the large $u$ limit of the relations \eqref{eq:RMRM}. Specifically, we consider the limit of $\frac{\eqref{eq:BA1}}{u^{2L-1}}$ and take the limit $u \to \infty$, arriving at
\beq 
\begin{split}
    \left[ \hat{\mathfrak{q}}_{\alpha}, \mathscr{A} (v, \alpha) \right] & = \lim_{u\to\infty} \frac{\mathscr{B} (v, \alpha)}{u^{2L-1}}  \left[ \left( \frac{\ri}{u} + \frac{\ri v +1}{u^2} + O \big( \frac{1}{u^3} \big) \right) (u^{2L+1} \mathscr{A}_{\infty}^{(1)} + u^{2L} \mathscr{A}_{\infty}^{(2)} + O (u^{2L-1})) + \right. \\
    & \left. \left( \frac{\ri}{u} + \frac{-\ri v}{u^2} + O \big( \frac{1}{u^3} \big) \right) (u^{2L+1} \mathscr{D}_{\infty}^{(1)} + u^{2L} \mathscr{D}_{\infty}^{(2)} + O (u^{2L-1}))  \right] \\
    & = \mathscr{B} (v, \alpha) \left( (2 \ri v +1 ) - 2 (2 \mathbf{S}^z + \alpha) \right) .
\end{split}
\label{eq:Binf1}
\eeq 

Similarly, we perform the same calculation for the limit of $\frac{\eqref{eq:BD1}}{v^{2L-1}}$ and take the limit $v \to \infty$, and we obtain
\beq 
    \left[ \mathscr{D} (u, \alpha) ,\hat{\mathfrak{q}}_{\alpha} \right]= \mathscr{B} (u, \alpha) \left( (3 - 2 \ri u ) - 2 (2 \mathbf{S}^z + \alpha) \right) .
\label{eq:Binf2}
\eeq 
From now on we focus on the iso-BAE case with $\beta=\alpha-1$. Let us consider an eigenstate $| \{ u_j \}_{j=1}^M  \rangle_{\alpha}$ of the transfer matrix
\beq 
    \mathbf{T}_{1/2} (u, \alpha, \alpha-1) | \{ u_j \}_{j=1}^M  \rangle_{\alpha} = T_{1/2}(u,\alpha,\alpha-1) | \{ u_j \}_{j=1}^M  \rangle_{\alpha} ,
\eeq 
We shall prove that the state $(\hat{\mathfrak{q}}_{\alpha})^p| \{ u_j \}_{j=1}^M  \rangle_{\alpha} $ with $p=L-2M+1$ is again an eigenstate of $\mathbf{T}_{1/2} (u, \alpha, \alpha-1)$, unless it is annihilated by $(\hat{\mathfrak{q}}_{\alpha})^p$. This can be proven as follows.\par

Assuming that $(\hat{\mathfrak{q}}_{\alpha})^p | \{ u_j \}_{j=1}^M \rangle_{\alpha}\ne 0$, by using \eqref{eq:Binf1} and \eqref{eq:Binf2} repeatedly, we obtain
\begin{align}
    \mathbf{T}_{1/2} (u , \alpha, \alpha-1)(\hat{\mathfrak{q}}_{\alpha})^p | \{ u_j \}_{j=1}^M \rangle_{\alpha} =(\hat{\mathfrak{q}}_{\alpha})^p \mathbf{T}_{1/2} (u , \alpha, \alpha-1) | \{ u_j \}_{j=1}^M \rangle_{\alpha}
       + \chi( u, \alpha ) (\hat{\mathfrak{q}}_{\alpha})^{p-1}\mathscr{B} (u , \alpha) | \{ u_j \}_{j=1}^M \rangle_{\alpha} ,
\end{align}
where the function $\chi(u,\alpha)$ is
\beq 
\begin{split}
    \chi (u , \alpha) & = \sum_{m=1}^p \left[ -(-u+\ri (\alpha-\tfrac{3}{2})) (2\ri u +1 - 2(L-2(M+p-n) + \alpha) ) \right. \\
    & \left. + (u+\ri (\alpha - \tfrac{1}{2})) (3 - 2 \ri u - 2(L-2(M+p-n) +\alpha ) ) \right]\,.
\end{split}
\eeq 
It is easy to check that when $p = L-2M+1$, we have $\chi(u,\alpha)=0$. In this case, we have
\begin{align}
\label{eq:commuteTQ}
[\mathbf{T}_{1/2} (u , \alpha, \alpha-1),(\hat{\mathfrak{q}}_{\alpha})^p ]| \{ u_j \}_{j=1}^M \rangle_{\alpha}=0\,.
\end{align}

For type-I states $\hat{\mathfrak{q}}_\alpha^{L-2M+1} | \psi \rangle^{\rm I} = 0 $. For type-II states $\hat{\mathfrak{q}}_\alpha^{L-2M+1} | \psi \rangle^{\rm II} \neq 0$ and is degenerate with $| \psi \rangle^{\rm II} $. Combing all magnon sectors, it is natural to define the operator 
\beq 
    \mathcal{Q}_\alpha = \sum_{M=0}^{[L/2]}\Pi_{L-M+1}\hat{\mathfrak{q}}_{\alpha}^{L-2M+1}\Pi_M ,
    \label{eq:constructQ}
\eeq 
and from \eqref{eq:commuteTQ}, we have
\beq 
    \left[ \mathcal{Q}_\alpha , \mathbf{T}_{1/2} (u , \alpha, \alpha-1) \right] = \left[ \mathcal{Q}_\alpha (q) , \mathbf{H}_\alpha \right] =  0 ,
\eeq 
which proves the statement in Sec. \ref{subsec:typeIandII}.

\subsection{XXZ model}
\label{app:ABAXXZ}

Now we perform a similar analysis for the XXZ model.
\paragraph{Algebraic Bethe ansatz} The $R$-matrix for the XXZ model is
\beq 
    \mathbf{R}_{ab} (t ) = t \bp q^{\sigma^z_b/2} & {\color{gray} 0 } \\ {\color{gray} 0 } & q^{-\sigma^z_b/2} \ep_a + (q - q^{-1}) \bp {\color{gray} 0 } & \sigma^-_b \\ \sigma^+_b & {\color{gray} 0 }  \ep_a + t^{-1} \bp -q^{-\sigma^z_b/2} & {\color{gray} 0 } \\ {\color{gray} 0 } & -q^{\sigma^z_b/2} \ep_a ,
\eeq 
where $q$ is the $q$-deformation parameter with the anisotropy parameter $\Delta = (q+q^{-1})/2$.

The boundary $K$-matrices are 
\beq 
\begin{split}
    & \mathbf{K}_{\mathrm{L} ,a} (t , \beta ) = t^{-1} \bp q^{\beta-1/2} & {\color{gray} 0 } \\ {\color{gray} 0 } & -q^{-\beta-1/2} \ep_a + t \bp -q^{-\beta+1/2} & {\color{gray} 0 } \\ {\color{gray} 0 } & q^{\beta+1/2} \ep_a, \\ 
    & \mathbf{K}_{\mathrm{R} ,a} (t , \alpha ) = t \bp q^{\alpha-1/2} & {\color{gray} 0 } \\ {\color{gray} 0 } & -q^{-\alpha-1/2} \ep_a + t^{-1} \bp -q^{-\alpha+1/2} & {\color{gray} 0 } \\ {\color{gray} 0 } & q^{\alpha+1/2} \ep_a .
\end{split}
\eeq 
The $R$- and $K$ matrices satisfy the YBE and boundary YBE respectively.\par

The monodromy matrix and the double-row transfer matrices are defined similarly as
\beq 
\begin{split}
    \mathbf{M}_a (t , \alpha , \beta) & =  \mathbf{K}_{\mathrm{L} ,a} (t , \beta )  \prod_{n=1}^L \mathbf{R}_{a n} (t q^{-1/2} ) \mathbf{K}_{\mathrm{R} ,a} (t , \alpha ) \prod_{n=L}^1 \mathbf{R}_{a n} (t q^{-1/2} ) \\ 
    & = \mathbf{K}_{\mathrm{L} ,a} (t , \beta )  \mathscr{M}_a(t,\alpha) \\
    & = \mathbf{K}_{\mathrm{L} ,a} (t , \beta )   \bp \mathscr{A} (t , \alpha) &  \mathscr{B} (t , \alpha ) \\  \mathscr{C} (t , \alpha ) &  \mathscr{D} (t , \alpha ) \ep_a ,
\end{split}
\eeq 
and
\beq 
    \mathbf{T}_{1/2} (t , \alpha, \beta ) = \mathrm{Tr}_a  \mathbf{M}_a (t , \alpha , \beta) .
\eeq 
The Hamiltonian can be obtained from the transfer matrix and is given by
\begin{align} 
    \mathbf{H}^{\rm XXZ}_{\alpha , \beta} = \sum_{n=1}^{L-1} \left[ \sigma^x_n \sigma^x_{n+1} + \sigma^y_n \sigma^y_{n+1} + \Delta (\sigma^z_n \sigma^z_{n+1} - 1 ) \right] + \frac{\sinh \eta}{\tanh (\alpha \eta)} (\sigma^z_L - 1) - \frac{\sinh \eta}{\tanh (\beta \eta)} (\sigma^z_1 - 1) 
    \label{eq:XXZHamiltonian}
\end{align}
where $\Delta=\cosh\eta=(q+q^{-1})/2$\,.

\paragraph{The iso-BAE flow} The iso-BAE flow requires that $\beta = \alpha-1$ in \eqref{eq:XXZHamiltonian}. Moreover, it is convenient to split into an undeformed Hamiltonian together with additional boundary terms. The undeformed Hamiltonian is obtained by taking following limit 
\beq 
    \mathbf{H}^{\rm XXZ}_0 = \lim_{\alpha \eta \to \infty} \mathbf{H}^{\rm XXZ}_{\alpha, \alpha-1}  = \sum_{n=1}^{L-1} \left[ \sigma^x_n \sigma^x_{n+1} + \sigma^y_n \sigma^y_{n+1} + \Delta (\sigma^z_n \sigma^z_{n+1} - 1 ) \right] + \sinh \eta (\sigma^z_L - \sigma^z_1) \,,
\eeq
while the iso-BAE XXZ Hamiltonian becomes
\beq 
    \mathbf{H}^{\rm XXZ}_{\alpha} = \mathbf{H}^{\rm XXZ}_{\alpha, \alpha-1} = \mathbf{H}^{\rm XXZ}_0 + \mathbf{V}_\alpha \, ,
\eeq 
where
\beq 
    \mathbf{V}^{\rm XXZ}_{\alpha} = \sinh \eta \left[ \big( \frac{1}{\tanh (\alpha \eta)} -1 \big) (\sigma^z_L - 1)  - \big( \frac{1}{\tanh ((\alpha-1)\eta)} -1 \big) (\sigma^z_1 - 1) \right] .
\eeq 

\paragraph{The operator $\mathcal{Q}_{\alpha}$} Now we focus on the iso-BAE case with $\beta=\alpha-1$. The operator $\mathcal{Q}_{\alpha}$ can be constructed similarly as in the XXX$_{1/2}$ case. The operator $\hat{\mathfrak{q}}_\alpha (q) $ is again proportional to the leading order of $\mathscr{B}(t,\alpha)$ operator in the $t \to \infty$ expansion, and we have
\beq 
    \hat{\mathfrak{q}}_\alpha (q) = \frac{1}{q-q^{-1}} \mathscr{B}_{\infty}^{(1)} = \sum_{m=1}^L  \left( - q^{-\alpha-1} \prod_{n=m+1}^L q^{-\sigma^z_n}  + q^{\alpha-1} \prod_{n=m+1}^L q^{\sigma^z_n} \right) \sigma^-_m .
\eeq 
The operator $\mathcal{Q}_{\alpha}$ is constructed exactly as in \eqref{eq:constructQ}. Following the same steps as in the previous subsection, we can prove that
\beq 
    \left[ \mathcal{Q}_\alpha (q) , \mathbf{T}_{1/2} (u , \alpha, \alpha-1) \right] = \left[ \mathcal{Q}_\alpha (q) , \mathbf{H}_\alpha \right] =  0 .
\eeq 
We would like to comment on the case when $q = \exp (\ri \pi \frac{\ell_1}{\ell_2})$ is a root of unity. In this case, we have
\beq 
    \left( \hat{\mathfrak{q}}_\alpha (q) \right)^{\ell_2} = 0 .
\eeq 
This implies that if $\ell_2$ is sufficiently small, we are not able to obtain all the type-II states with $M > \lfloor \frac{L}{2} \rfloor$ using $\mathcal{Q}_\alpha (q)$ anymore. In the meantime, when $q$ is at root of unity, additional degeneracies appear in the spectrum of $\mathbf{H}^{\rm XXZ}_0$ due to the special properties of the representation theory of $U_q (\mathfrak{sl}_2)$ at root of unity. All of these subtleties have to be taken in account and we postpone the discussion for the root of unity case to a later work.

\subsection{\texorpdfstring{$\mathrm{XXX}_s$}{XXXs} model}
\label{app:ABAXXXs}

Finally, we consider the higher spin $\mathrm{XXX}_s$ model. 
\paragraph{Algebraic Bethe ansatz} We take the auxiliary space to be spin-$1/2$, which is sufficient for our purpose.
The Lax operator in this case reads
\beq 
    \mathbf{L}_{a b} (u) = u \bp \mathbb{I}_b &  {\color{gray} 0 } \\  {\color{gray} 0 }  & \mathbb{I}_b \ep_a + \ri \bp \mathbf{S}^z_b &  \mathbf{S}^-_b \\  \mathbf{S}^+_b & - \mathbf{S}^z_b \ep_a ,
\eeq 
where $\mathbf{S}^\alpha$ are spin-$s$ generators of the SU(2) algebra. The boundary $K$-matrices are the same as the spin-1/2 case, \emph{i.e.} \eqref{eq:boundaryKXXX}. The monodromy matrix becomes
\beq 
\begin{split}
    \mathbf{M}_a (u, \alpha , \beta) & =  \mathbf{K}_{\mathrm{L} ,a} (u , \beta )  \prod_{n=1}^L \mathbf{L}_{a n} (u-\ri/2 ) \mathbf{K}_{\mathrm{R} ,a} (u , \alpha ) \prod_{n=L}^1 \mathbf{L}_{a n} (u-\ri/2 ) \\ 
    & = \mathbf{K}_{\mathrm{L} ,a} (u , \beta )  \mathscr{M}_a(u,\alpha) \\
    & = \mathbf{K}_{\mathrm{L} ,a} (u , \beta )   \bp \mathscr{A} (u , \alpha) &  \mathscr{B} (u , \alpha ) \\  \mathscr{C} (u , \alpha ) &  \mathscr{D} (u , \alpha ) \ep_a ,
\end{split}
\eeq 
and the double-row transfer matrix reads
\beq 
    \mathbf{T}_{1/2} (u, \alpha, \beta ) = \mathrm{Tr}_a  \mathbf{M}_a (u, \alpha , \beta) = (-u+\ri (\beta - \tfrac{1}{2})) \mathscr{A} (u , \alpha ) + (u+\ri (\beta + \tfrac{1}{2})) \mathscr{D} (u , \alpha ) .
\eeq 
We again have the same relations between $\mathscr{M}_a (u, \alpha)$ as in \eqref{eq:RMRM}, since the intertwiner remains the same. Hence, the algebra between the quantum operators $\mathscr{A}$, $\mathscr{B}$, $\mathscr{C}$ and $\mathscr{D}$ remain the same as the spin-1/2 case too.

Note that in the higher spin XXX$_s$ model, if we want to construct the eigenstate and find the BAE, it is sufficient to consider the transfer matrix with spin-$1/2$ representation in the auxiliary space. However, if we want to obtain the Hamiltonian, we have to take the auxiliary space to be in the same representation of the local quantum space. These transfer matrices can be obtained by the fusion procedure~\cite{Kulish:1981gi}. For example, for $s=1$, the Hamiltonian is given by
\beq 
\label{eq:Hhighspin}
\begin{split} 
\mathbf{H}^{(1)}_{\alpha, \beta} & = \sum_{n=1}^{L-1} \left[ \vec{\mathbf{S}}_n \cdot \vec{\mathbf{S}}_{n+1} - \big( \vec{\mathbf{S}}_n \cdot \vec{\mathbf{S}}_{n+1}  \big)^2  \right] \\
    & + \left( \frac{1}{\alpha+1/2} + \frac{1}{\alpha-1/2} \right) \mathbf{S}^z_L + \left( \frac{1}{\alpha+1/2} - \frac{1}{\alpha-1/2} \right) \big( \mathbf{S}^z_L \big)^2 \\
    & - \left( \frac{1}{\beta+1/2} + \frac{1}{\beta-1/2} \right) \mathbf{S}^z_1 + \left( \frac{1}{\beta+1/2} - \frac{1}{\beta-1/2} \right) \big( \mathbf{S}^z_1 \big)^2 + 2\left( \frac{1}{\beta-1/2} - \frac{1}{\alpha+1/2} \right) ,
\end{split}
\eeq 
where the spin-1 operators are defined as 
\beq 
    \mathbf{S}^z_m = \mathbb{I}^{\otimes (m-1)} \otimes \bp 1 & 0&0 \\0 & 0 &0 \\
   0 &0 & -1 \ep \otimes \mathbb{I}^{\otimes (L-m)}  , \quad \mathbf{S}^+_m = \mathbb{I}^{\otimes (m-1)} \otimes \bp 0 & \sqrt{2} &0 \\ 0& 0 & \sqrt{2} \\
    0&0 &0 \ep \otimes \mathbb{I}^{\otimes (L-m)} = \left( \mathbf{S}^-_m \right)^\dag .
    \label{eq:spin1operators}
\eeq 
The local Hilbert space is $\mathbb{C}^3$ and we choose the three basis states as 
\begin{align}
|+\rangle=\left(\begin{array}{c}1\\0\\0 \end{array}\right),\qquad 
|0\rangle=\left(\begin{array}{c}0\\1\\0 \end{array}\right),\qquad
|-\rangle=\left(\begin{array}{c}0\\0\\1 \end{array}\right)\,.
\end{align}

The Hamiltonian for spin-$s$ model can be calculated by the fused higher spin transfer matrices. For more details we refer to \cite{Kulish:1981gi, Babujian:1983ae}.

\paragraph{The iso-BAE flow} For higher spin cases, the iso-BAE flow again occurs at $\beta = \alpha-1$. We demonstrate on the spin-1 Hamiltonian \eqref{eq:Hhighspin}. 

The spin-1 iso-BAE Hamiltonian is obtained by taking $\beta=\alpha-1$ in \eqref{eq:Hhighspin}, \emph{i.e.}
\beq 
    \mathbf{H}^{(1)}_{\alpha} = \mathbf{H}^{(1)}_{0} + \mathbf{V}^{(1)}_\alpha , 
\eeq 
where the Hamiltonian with SU(2) symmetry is
\beq 
\mathbf{H}^{(1)}_{0} = \lim_{\alpha \to \infty} \mathbf{H}^{(1)}_{\alpha} =  \sum_{n=1}^{L-1} \left[ \vec{\mathbf{S}}_n \cdot \vec{\mathbf{S}}_{n+1} - \big( \vec{\mathbf{S}}_n \cdot \vec{\mathbf{S}}_{n+1}  \big)^2  \right] ,
\eeq 
and the deformation term is
\beq 
\begin{split}
    \mathbf{V}^{(1)}_{\alpha} & = \left( \frac{1}{\alpha+1/2} + \frac{1}{\alpha-1/2} \right) \mathbf{S}^z_L + \left( \frac{1}{\alpha+1/2} - \frac{1}{\alpha-1/2} \right) \big( \mathbf{S}^z_L \big)^2 \\
    & - \left( \frac{1}{\alpha-1/2} + \frac{1}{\alpha-3/2} \right) \mathbf{S}^z_1 + \left( \frac{1}{\alpha-1/2} - \frac{1}{\alpha-3/2} \right) \big( \mathbf{S}^z_1 \big)^2 + 2\left( \frac{1}{\alpha-3/2} - \frac{1}{\alpha+1/2} \right) .
\end{split}
\label{eq:spin1Valpha}
\eeq 
The Hamiltonian mentioned in Sec. \ref{subsec:higherspinXXXs} is given above.

\paragraph{The $\mathcal{Q}_{\alpha}^{(s)}$ operator} The $\mathcal{Q}_{\alpha}^{(s)}$ operator is constructed in the same way as before. The operator $\hat{\mathfrak{q}}_\alpha^{(s)}$ is proportional to the leading order of the $\mathscr{B}(u)$ operator in the $u \to \infty$ limit and is given by
\beq 
     \hat{\mathfrak{q}}_\alpha^{(s)} = -2 \alpha \sum_{m=1}^L \mathbf{S}^-_m - 4 \sum_{m<n} \mathbf{S}^-_m \mathbf{S}^z_n - \sum_{m=1}^L \{ \mathbf{S}^-_m , \mathbf{S}^z_m \} \,.
\eeq 
Here the only difference between different $s$ is the explicit representation for the spin operators. The $\mathcal{Q}_{\alpha}^{(s)}$ operator is constructed as
\beq 
\mathcal{Q}_{\alpha}^{(s)} =\sum_{M=0}^{\lfloor sL \rfloor }\Pi_{2 s L-M+1}(\hat{\mathfrak{q}}^{(s)}_{\alpha})^{2 s L-2M+1} \Pi_M\,,
\eeq 
The proof of commutativity of $\mathcal{Q}_{\alpha}^{(s)}$ with $\mathbf{T}_{1/2}(u,\alpha,\alpha-1)$ follows similarly as in the previous subsections. In order to prove the commutativity of $\mathcal{Q}_{\alpha}^{(s)}$ with $\mathbf{H}^{(s)}_{\alpha}$, we simply prove that $\mathcal{Q}_{\alpha}^{(s)}$ commutes with higher transfer matrices, which follows directly from the fusion procedure~\cite{Kulish:1981gi}.

\section{\texorpdfstring{$TQ$}{TQ}-relation and rational \texorpdfstring{$Q$}{Q}-system}
\label{app:QsystemXXX}
The spectrum of integrable spin chains is encoded in the solutions of BAE. However, working directly with BAE has a number of drawbacks. First, there are solutions of BAE whose corresponding Bethe states are not eigenstates of the transfer matrix. These solutions are called unphysical solutions. Second, as we have seen in the main text, in the iso-BAE flow the naive BAE do not give the complete physical states. To solve these problems, it is better to apply the $TQ$-relation or the rational $Q$-system.\par

The $TQ$-relation for open spin chains have been known for some time \cite{Sklyanin:1988yz} (see \cite{Wang_2015} for detailed discussions and more references). It can be derived either from ABA or by an explicit construction of Baxter's $Q$-operator \cite{Frassek:2015mra,Vlaar2020,Frassek:2022izn}. The $TQ$-relation is sufficient to describe all physical states, which solves the second drawback mentioned above. However, $TQ$-relation still has unphysical solutions. To solve this problem, we exploit the rational $Q$-system, which contain all and only physical solutions. This powerful new approach was developed only recently \cite{Bajnok:2019zub,Nepomechie:2019gqt} for open chains. In this section, we review the $TQ$-relation and the rational $Q$-system for the three type of iso-BAE flows in the main text.  The rational $Q$-system for XXX$_s$ model with open boundary conditions have not been worked out in the literature. It is constructed in this section, which also constitutes a new result of this work. We also present concrete examples which demonstrate the structure of the Hilbert space based on the solutions of rational $Q$-systems.

\subsection{The XXX\texorpdfstring{$_{1/2}$}{1/2} model}
The $TQ$-relation of the open XXX$_{1/2}$ reads
\beq
    - u T_{1/2} (u) Q (u) =  u^+ T_0^+ (u) g^- (u) Q^{2-}  (u) +  u^- T_0^-  (u) f^+ (u) Q^{2+}  (u) ,
    \label{eq:TQ}
\eeq 
where $T_0 (u) = u^{2L}$ and $Q(u)$ is a polynomial whose zeros are the Bethe roots,
\beq 
    Q(u) = \prod_{m=1}^M (u - u_m)(u + u_m)\,.
\eeq 
$T_{1/2} (u)$ is the eigenvalue of the transfer matrix defined in Sec. \ref{app:ABA}, which is also a polynomial in $u$. The BAE can be obtained from $TQ$-relation \eqref{eq:TQ} by taking $u=u_m$. Therefore, solutions of the $TQ$ relation also solve BAE.

\paragraph{Rational $Q$-system}  As mentioned above, solutions to the $TQ$-relation contain unphysical solutions \cite{Marboe:2016yyn, Bajnok:2019zub, Nepomechie:2019gqt, Hou:2023jkr}. Therefore, we need additional conditions to restrict the number of solutions. One nice way to achieve this is the rational $Q$-system \cite{Marboe:2016yyn, Bajnok:2019zub,Nepomechie:2019gqt}.
The rational Q-system for the spin-$1/2$ XXX spin chain with open boundaries \cite{Nepomechie:2019gqt} is defined by the following set of functional equations called $QQ$-relations
\beq 
\begin{split}
    & u Q_{1,n} = f^{(n-1)-} Q_{1,n-1}^+ - g^{(n-1)+} Q_{1,n-1}^- ,\quad n=0,1,\ldots,M \\
    & u Q_{0,n} Q_{1,n-1} =  Q_{0,n-1}^+ Q_{1,n}^- -  Q_{0,n-1}^- Q_{1,n}^+,\quad n=0,1,\ldots,L-M,
\end{split}
\label{eq:Qsystems}
\eeq 
where $Q_{1,0}(u)$ is identified with Baxter's $Q$-polynomial in the $TQ$-relation. We require that all the functions $Q_{a,s}(u)$ are polynomials in $u$ and solve the $QQ$-relations \eqref{eq:Qsystems} with the boundary conditions
\begin{align}
    Q_{0,0}(u) = u^{2L}, \qquad
    Q_{1,0}(u) = \prod_{m=1}^M (u - u_m)(u + u_m) = \prod_{m=1}^M (u^2 - u_m^2) .
\end{align}
We find that even though the BAE \eqref{eq:BAE} for the iso-BAE systems with different values of $\alpha$ are identical, the rational $Q$-systems differ. Therefore the $Q$-system indeed describe the full spectrum for the iso-BAE flow. By solving the $Q$-system, we indeed obtain two kinds of solutions. The solutions associated with type-I states remain the same for the iso-BAE systems with different values of $\alpha$, while the solutions associated with type-II states differ.

\paragraph{The undeformed model} The undeformed model $\mathbf{H}_0$ in \eqref{eq:H0XXX} is obtained in the limit $\alpha,\beta\to\infty$ and is simply the open XXX spin chain with free boundary conditions at both ends. In terms of the $TQ$-relation or the rational $Q$-system, this limit corresponds to taking $f(u)=g(u)=1$. This model has the global SU(2) symmetry. The highest weight states correspond to the type-I states in the deformed model.

\begin{figure}[h!]
    \centering
    \includegraphics[width=.55\linewidth]{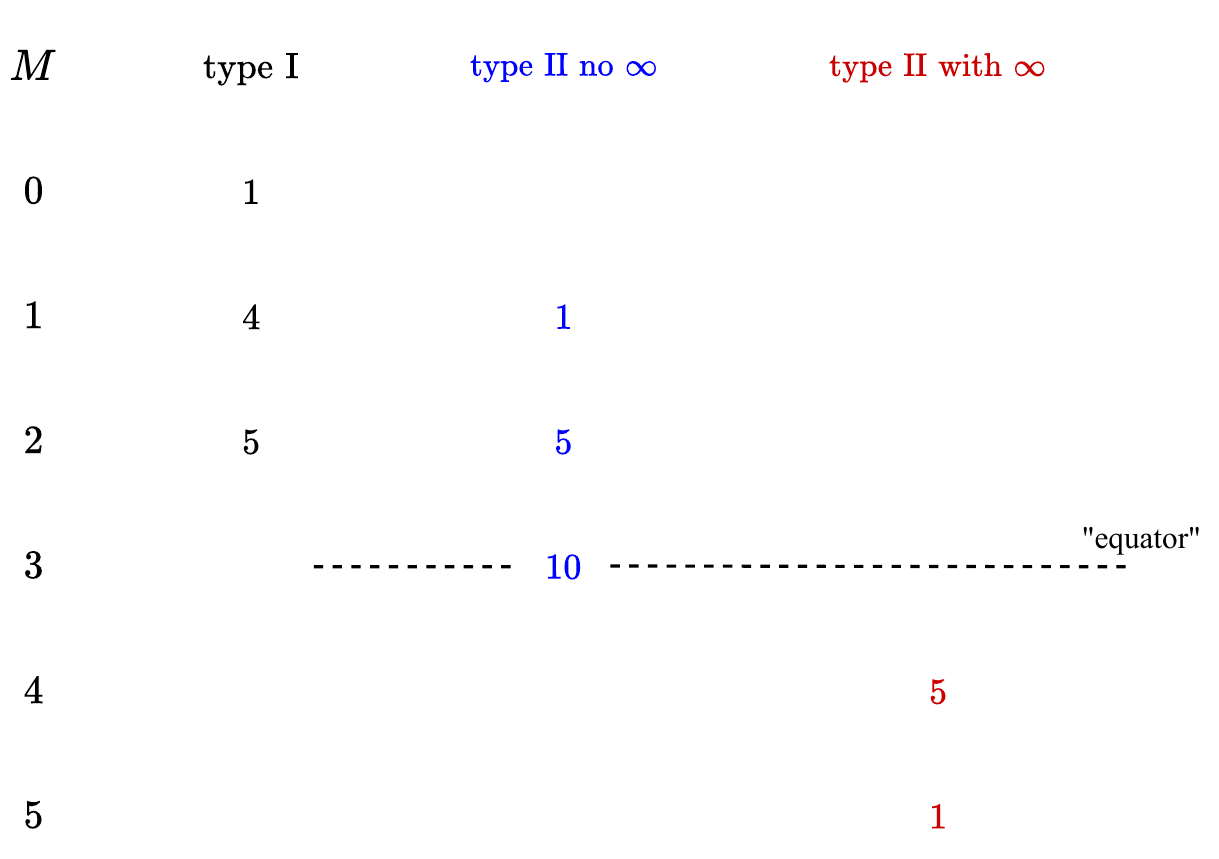}
    \caption{The Hilbert space structure for ${\rm XXX}_{1/2}$ with $L=5$ for iso-BAE systems with arbitrary boundary parameter $\alpha$. The doubly degeneracies between type-II eigenstates of $M=1\leftrightarrow M=5$, and $M=2\leftrightarrow M=4$, respectively, can be observed clearly.}
    \label{fig:Hilbertspace1}
\end{figure}

\begin{figure}[h!]
    \centering
    \includegraphics[width=.55\linewidth]{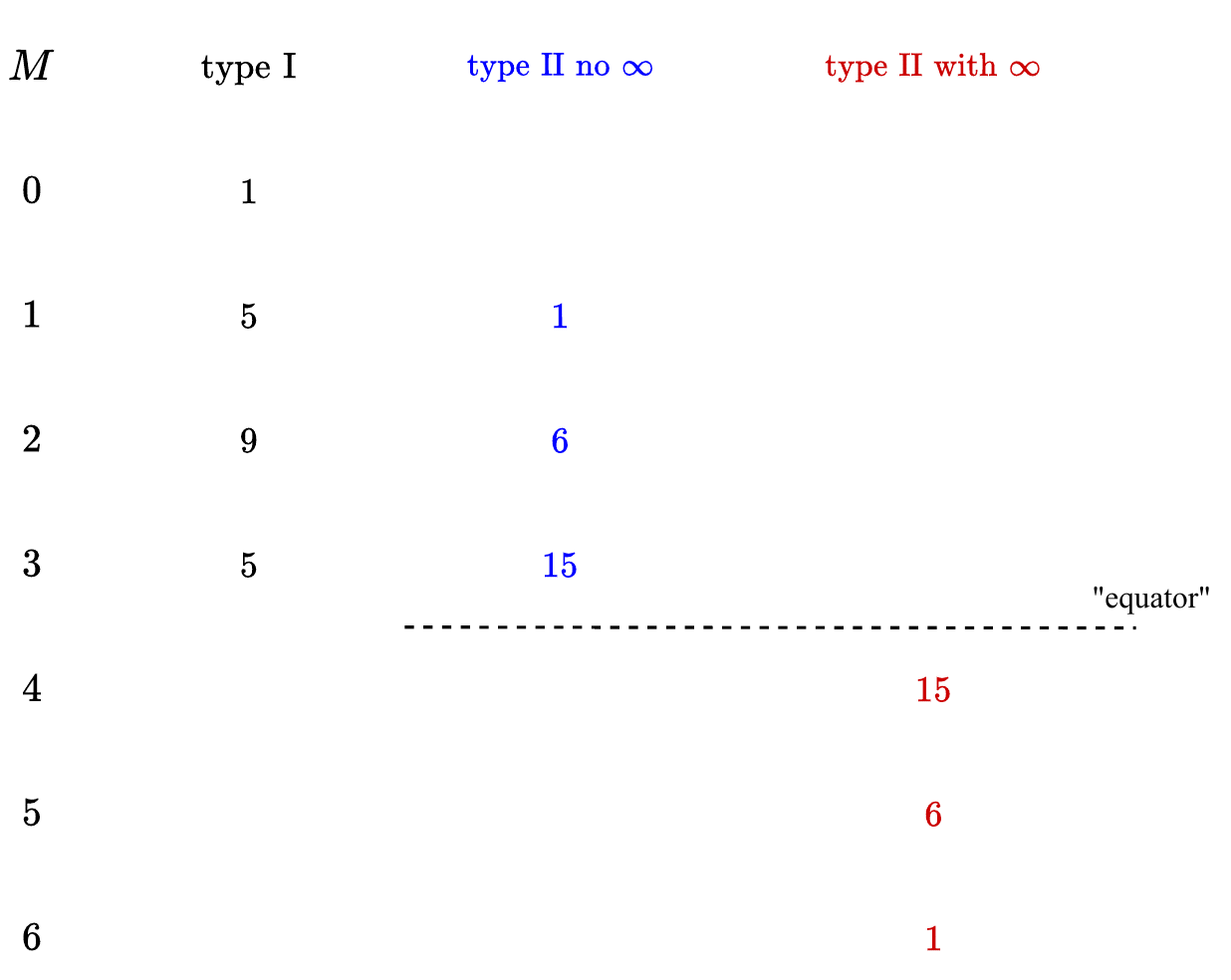}
    \caption{The Hilbert space structure for ${\rm XXX}_{1/2}$ with $L=6$ for iso-BAE systems with arbitrary boundary parameter $\alpha$. The doubly degeneracies between type-II eigenstates of $M=1\leftrightarrow M=6$, $M=2 \leftrightarrow M=5$, and $M=3\leftrightarrow M=4$, respectively, can be observed clearly.}
    \label{fig:Hilbertspace2}
\end{figure}

\paragraph{Explicit examples} In order to demonstrate the structure of Hilbert space for even and odd system size $L$ in Sec. \ref{subsec:structureHilbert}, we give two explicit examples for the iso-BAE systems of $L=6$ and $L=5$ with finite value of $\alpha$. From Fig. \ref{fig:Hilbertspace1} and Fig. \ref{fig:Hilbertspace2}, we observe the difference between even and odd system size. All the type-II states are doubly degenerate for the even $L$. For odd $L$,  type-II states are doubly degenerate except for $M = \frac{L+1}{2}$. Analogous to the model with U(1) and spin-flip $\mathbb{Z}_2$ symmetry, we recognize an effective ``equator'' for type-II state at $M = \frac{L+1}{2}$, hinting at an effective $\mathbb{Z}_2$ symmetry.

\subsection{XXZ model}
\label{app:QsystemXXZ}
The $TQ$-relation for the XXZ spin chain becomes
\begin{align}
\label{eq:TQXXZ}
-\sinh(2u)T_{1/2}(u)Q(u)=&\,\sinh(2u+\eta)\sinh^{2L}(u+\tfrac{\eta}{2})g^-(u)Q^{--}(u)\\\nonumber
&\,+\sinh(2u-\eta)\sinh^{2L}(u-\tfrac{\eta}{2})f^+(u)Q^{++}(u)
\end{align}
with
\begin{align}
Q(u)=\prod_{k=1}^M\sinh(u-u_k)\sinh(u+u_k)\,.
\end{align}
and $T_{1/2}(u)$ being the eigenvalue of the transfer matrix. Taking $u=u_m$ in the $TQ$-relation \eqref{eq:TQXXZ}, we obtain the BAE
\beq 
    \frac{g(u_m - \eta /2)}{f(u_m + \eta /2)} \left( \frac{\sinh(u_m+\eta/2)}{\sinh(u_m-\eta/2)} \right)^{2L} \prod_{k\neq m}^M S^{\rm XXZ} (u_m, u_k) = 1 ,
\label{eq:BAE-XXZ}
\eeq 
where the scattering phase becomes
\beq 
    S^{\rm XXZ} (u_m, u_k) = \frac{\sinh(u_m - u_k -\eta)}{\sinh(u_m - u_k+\eta)} \frac{\sinh(u_m + u_k -\eta)}{\sinh(u_m + u_k+\eta)} .
\eeq 
The boundary polynomials are
\beq 
    f(u) = g(-u) = \sinh (u - \alpha \eta) \sinh (u+ \beta \eta)  .
\eeq 
We have introduced the shorthand notation $F^{\pm}(u)=F(u\pm\tfrac{\eta}{2})$. In the trigonometric case, sometimes it is more convenient to work with the multiplicative variables $t=e^u$. In terms of multiplicative variables we have  $F^{n\pm} (t) = F( t q^{\pm n/2} ) $ and with a slight abuse of notation
\begin{align}
f (t) = g (t^{-1}) =\frac{1}{4}(t q^{-\alpha} -t^{-1} q^{\alpha} )(t q^{\beta} -t^{-1} q^{-\beta} ).
\end{align}

\paragraph{Rational $Q$-system} The rational $Q$-system for the XXZ model with open boundaries is similar to the $Q$-system for the XXX model. The main difference is that $Q$-polynomials are now Laurent polynomials in $t$ instead of polynomials in $u$. With $q = e^\eta$, the $QQ$-relations for the XXZ spin chain become \cite{Nepomechie:2019gqt}
\beq 
\begin{split}
    & (t^2- t^{-2}) Q_{1,n} = f^{(n-1)-} Q_{1,n-1}^+ - g^{(n-1)+} Q_{1,n-1}^- , \quad n=1,\ldots,M,\\
    & (t^2- t^{-2}) Q_{0,n} Q_{1,n-1} =  Q_{0,n-1}^+ Q_{1,n}^- -  Q_{0,n-1}^- Q_{1,n}^+ ,\quad n=1,\ldots,L-M\,.
\end{split}
\eeq 
To solve the $Q$-system, we require that the functions $Q_{a,s}(t)$ to be Laurent polynomials,
\beq 
    Q_{a,s} (t) = \sum_{n=0}^L c^{(a,s)}_n (t^{2n} + t^{-2n} )
\eeq 
with the following boundary condition
\beq 
\begin{split}
    & Q_{0,0}(t) = \frac{1}{2^{2L}}\left( t  - t^{-1} \right)^{2L}, \\
    & Q_{1,0}(t) = Q(t) = \frac{1}{2^{2M}}\prod_{m=1}^M \left( t t_m^{-1} - t^{-1} t_m \right)\,. \\
    \end{split}
\eeq 
The iso-BAE flow is at $\beta=\alpha-1$ where $f^+=g^-$ and they cancel out in \eqref{eq:BAE-XXZ}. In this case, the BAE \eqref{eq:BAE-XXZ} describe the type-I states.

The type-II states correspond to the solutions with a fixed root $v_M=\pm(\alpha-\tfrac{1}{2})\eta$ and the rest $(M-1)$ roots being the solutions of the following reduced BAE
\begin{align}
\frac{\tilde{g}(v_m-\tfrac{\eta}{2})}{\tilde{f}(v_m+\tfrac{\eta}{2})}\left(\frac{\sinh(v_m+\tfrac{\eta}{2})}{\sinh(v_m-\tfrac{\eta}{2})}\right)^{2L}\prod_{j\ne m}^{M-1}S^{\text{XXZ}}(v_m,v_j)=1,\qquad m=1,\ldots,M-1\,,
\end{align}
where
\begin{align}
\tilde{f}(u)=\tilde{g}(-u)=\sinh(u+\alpha\eta)\sinh(u-(\alpha-1)\eta)\,.
\end{align}
We again have an effective BAE with $\tilde{\alpha} = -\alpha$ and $\tilde{\beta} = - \beta = - (\alpha-1)$.

The structure of Hilbert space in the XXZ case is the same as the XXX case described above, when $q$ is not a root of unity. We shall not repeat the results here.

\subsection{The \texorpdfstring{$\mathrm{XXX}_s$}{XXXs} model}
\label{app:Qsystemhigherspin}

For the higher spin XXX$_s$ model, the $TQ$-relation becomes 
\beq 
    - u T_{1/2} Q = u^+ T_0^{(2s)+} g^-  Q^{2-} + u^- T_0^{(2s)+} f^+ Q^{2+} ,
\eeq 
where $T_0 = u^{2L}$. The $Q(u)$, $f(u)$ and $g(u)$ polynomials are the same for the spin-1/2 case.
The BAE are obtained by taking $u=u_m$ in the $TQ$-relation above, leading to
\beq 
    \frac{g(u_m - \ri /2)}{f(u_m + \ri /2)} \left( \frac{u_m + 2 s \ri}{u_m - 2 s \ri} \right)^L \prod_{k\neq m}^M S(u_m, u_k) = 1 .
\eeq 
The scattering phase $S(u,v)$ is the same as the $s=1/2$ case, cf. \eqref{eq:scatteringphase}.

\paragraph{Rational $Q$-system} We can generalize the rational $Q$-system for spin-$1/2$ XXX model with open boundaries to its higher spin counterpart, using the method developed in \cite{Hou:2023ndn}. The difference is that the boundary conditions for the finite difference equations are different. We would like to remark that even though the rational $Q$-systems with open boundaries for $\mathrm{XXX}_s$ models is a simple generalization to the periodic case, the explicit formulae have not appeared in the literature as far as we know. Explicitly, the $QQ$-relations read
\beq 
\begin{split}
    & u Q_{1,n} = f^{(n-1)-} Q_{1,n-1}^+ - g^{(n-1)+} Q_{1,n-1}^- ,\quad n=0,\ldots,M \\
    & u Q_{0,n} Q_{1,n-1} =  Q_{0,n-1}^+ Q_{1,n}^- -  Q_{0,n-1}^- Q_{1,n}^+ ,\quad n=0,\ldots,2sL-M\,.
\end{split}
\eeq 
We solve the $QQ$-relations by requiring $\frac{1}{\prod_{k=1}^{2s-1} (u-\ri (s-k))^L} Q_{0,1}$ and the rest of $Q_{a,s}(u)$ functions are polynomials in $u$ \cite{Hou:2023ndn} with the following boundary conditions 
\begin{align} 
    Q_{0,0}(u)=&\, \prod_{k=0}^{2s-1} \left( u - \big( s-k-\frac{1}{2} \big)\ri\right)^L \left( u +\big( s-k-\frac{1}{2} \big)\ri\right)^L\,, \\\nonumber
    Q_{1,0}(u)=&\, Q(u) = \prod_{m=1}^M  (u - u_m) (u+u_m) .
\end{align}

\paragraph{Structure of the Hilbert space} As an explicit example, we focus on the spin-1 model. Higher spin cases are analogous. For $\mathrm{XXX}_1$ model, we have
\beq 
    Q_{0,0}(u) = \left( u +\frac{\ri}{2} \right)^{2L} \left( u -\frac{\ri}{2} \right)^{2L} = (u^+)^{2L} (u^-)^{2L} .
\eeq

The iso-BAE condition is the same $\beta=\alpha-1$ as before. The iso-BAE spin-1 Hamiltonian can be found in \eqref{eq:Hhighspin}.

The ${\rm XXX}_1$ iso-BAE spectrum again splits into two parts. The type-I states are invariant under the change of parameter $\alpha$; type-II states have a fixed Bethe root at $v_M = \ri (\alpha - \frac{1}{2})$ and the energies change with $\alpha$.\par 

From the numerical solutions to the rational $Q$-system, we find that the kernel operator $\mathcal{Q}_\alpha^{(1)}$ is spanned by the type-I states. In addition, similar to the spin-$1/2$ case, the type-II states are doubly degenerate. Specifically, for a type-II state with $M \leq L$, we have
\beq 
    \mathcal{Q}_\alpha^{(1)} |\{v_j \}_{j=1}^M  \rangle_{\alpha}^{\rm II} =|\{v_j \}_{j=1}^M \cup \{ \infty \times p^\prime \} \rangle_{\alpha}^{\rm II}
\eeq 
is a degenerate eigenstate of the transfer matrix (and Hamiltonian) with $|\{v_j \}_{j=1}^M  \rangle_{\alpha}^{\rm II} $.

\begin{figure}[h!]
    \centering
    \includegraphics[width=.55\linewidth]{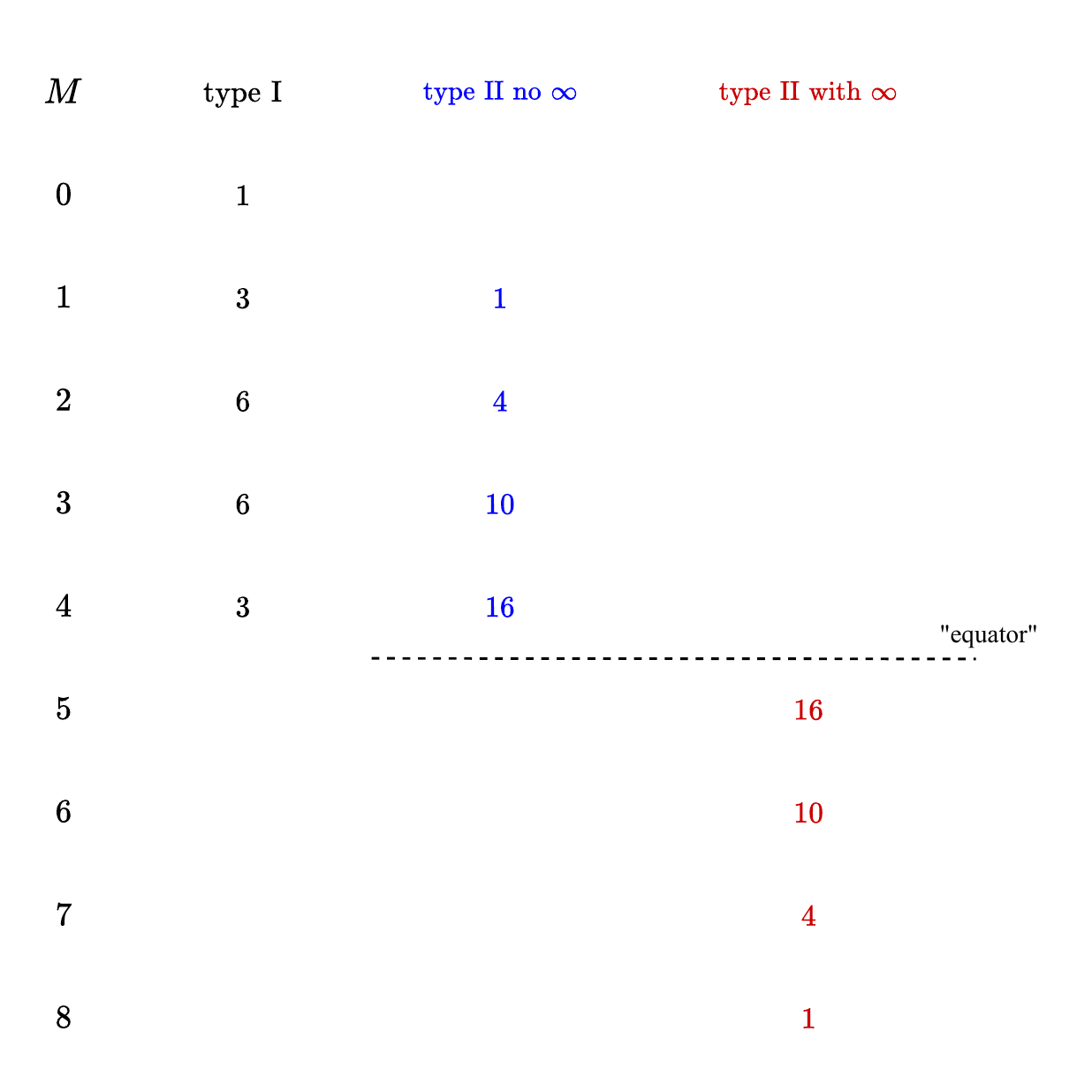}
    \caption{The Hilbert space structure for $L=4$ for spin-1 iso-BAE systems with arbitrary boundary parameter $\alpha$. The doubly degeneracies between type-II eigenstates can be observed clearly.}
    \label{fig:spin1Hilbert}
\end{figure}

The Hilbert space decomposes into type-I and type-II states in the spin-1 case. This can be shown by the following counting.
First of all, from combinatorics, the total number of states with $M$ magnons on the ferromagnetic state $| + + \cdots + \rangle$ is
\beq 
    \mathcal{N}^{(1)} (L,M) = \sum_{j=0}^{\lfloor M /2 \rfloor} \bp L \\ j \ep \bp L - j \\ M - 2j \ep = \mathcal{N}^{(1)} (L,2L-M) , \quad M \leq L
\eeq 
where $j$ is the number of $|- \rangle$ and $(M-2j)$ is the number of $|0 \rangle$ in a state with $M $ magnons in total. It is straightforward to verify that
\beq 
    \sum_{M=0}^{2L} \mathcal{N}^{(1)} (L,M)  = 3^L .
\eeq 
From BAE analysis, the number of type-I states is equal to the number of highest weight states for the undeformed model, \emph{i.e.}
\beq 
    \mathcal{N}_{\text{I}}^{(1)} (L,M)= \mathcal{N}^{(1)} (L,M) - \mathcal{N}^{(1)} (L,M-1) , \quad M \leq L .
\eeq 
The number of type-II states for $M\leq L$ is
\beq 
    \mathcal{N}_{\text{II}}^{(1)} (L,M) = \mathcal{N}^{(1)} (L,M) - \mathcal{N}^{(1)}_{\rm I} (L,M) = \mathcal{N}^{(1)} (L,M-1) , \quad 1 \leq M \leq L ,
\eeq 
and $\mathcal{N}_{\text{II}}^{(1)} (L,0) = 0$.Since there's no type-I states for $M>L$, we have
\beq 
    \mathcal{N}_{\text{II}}^{(1)} (L,M) = \mathcal{N}^{(1)} (L,M) =  \mathcal{N}^{(1)} (L,2L-M) , \quad M > L. 
\eeq 
All the counts are also supported by numerical checks. We present an example of the number of type-I and type-II states for the spin-1 iso-BAE systems with $L=4$ in Fig. \ref{fig:spin1Hilbert}.

\section{\texorpdfstring{$\alpha \to 1^+$}{\textalpha  to 1+} limit}
\label{app:alpha1limit}

When we take the limit $\alpha \to 1^+$ (or effectively $\alpha \to 0^-$, which is analogous and we shall not expand the discussion), one of the boundary magnetic fields diverges.

This implies that we can separate the spectrum into the finite-valued part and the infinite-valued part. Since the spectrum of the type-I states remain intact, they should belong to the finite-valued part of the spectrum.

Effectively, we can decouple the spin at the first site, since the boundary magnetic field $\frac{-1}{\alpha-1} \sigma^z_1$ in \eqref{eq:Hamiltonian} diverges. The finite-valued part of the spectrum consists of the eigenstates in the form of 
\beq 
    |\!\uparrow \rangle_1 \otimes | \psi \rangle_{2,3, \cdots L} .
\eeq
After freezing the first spin, we obtain an effective Hamiltonian acting on $2,3, \cdots L$ sites with an effective boundary magnetic field on the second site. More specifically, the state $| \psi \rangle_{2,3, \cdots L}$ is hence an eigenstate of the effective Hamiltonian,
\beq 
    \mathbf{H}_{\rm eff} = \sum_{n=2}^{L-1} \left( \vec{\sigma}_n \cdot \vec{\sigma}_{n+1} -1 \right) + ( \sigma^z_L -1 ) + ( \sigma^z_1 -1 ) ,
    \label{eq:Heff}
\eeq 
which is an open ${\rm XXX}_{1/2}$ model with $(L-1)$ sites and $\alpha_{\rm eff}=-\beta_{\rm eff} = 1$, cf. the footnote in Sec. \ref{sec:BAE} of the main text.

Since the effective Hamiltonian \eqref{eq:Heff} has only $(L-1)$ sites, the BAE again coincide with the iso-BAE case, but with a different mechanism,
\beq 
    \frac{g(u_m-\ri/2)}{f(u_m+\ri/2)}\left( \frac{u_m + \ri/2 }{u_m - \ri /2} \right)^{2L-2} \prod_{k\neq m}^M S(u_m, u_k) = 1 \,\, \Rightarrow \,\, \left( \frac{u_m + \ri/2 }{u_m - \ri /2} \right)^{2L} \prod_{k\neq m}^M S(u_m, u_k) = 1 ,
\eeq
where $f(u) = (u-\ri)^2 = g(-u)$ in this case.

Similarly, we can use the rational $Q$-system to study the spectrum of the effective Hamiltonian. We conclude that the effective Hamiltonian \eqref{eq:Heff} with $(L-1)$ sites again contains the type-I solutions of the iso-BAE systems (with $L$ sites). The same observation in the $q$-deformed case has been used in \cite{Morin-Duchesne:2015afa} to prove the reality of the spectrum of the quantum-group-invariant XXZ spin chain in the easy-plane regime.

\section{Ground states along the iso-BAE flow}
\label{app:phases}
\begin{figure}[h!] 
\centering
\includegraphics[scale=0.5]{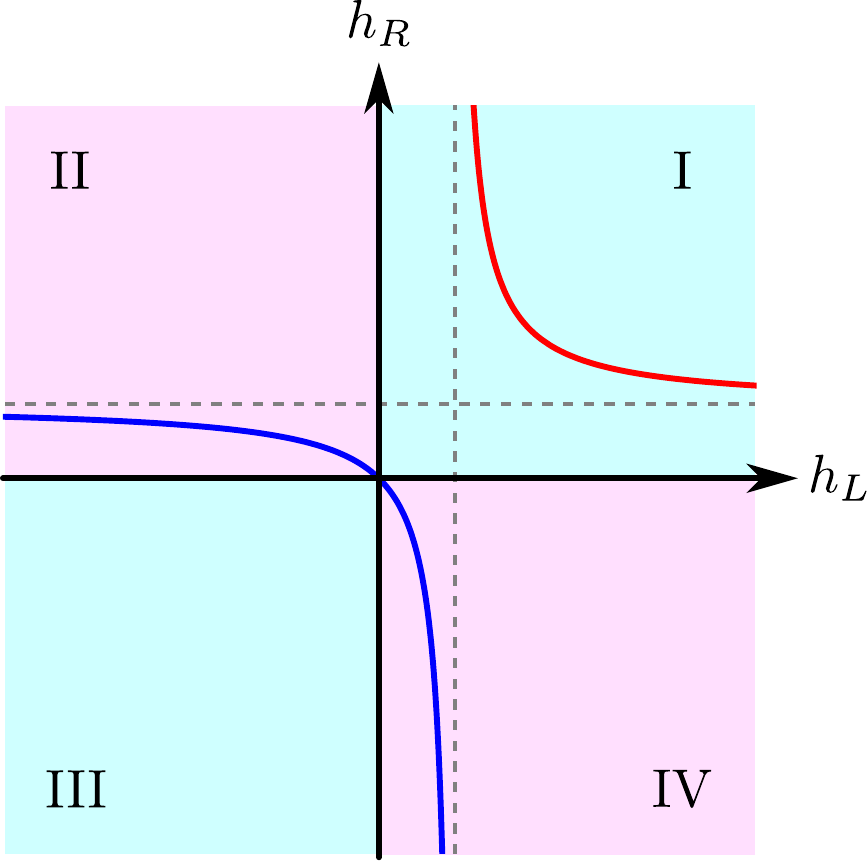}
\caption{Various ground states with respect to the boundary magnetic field $h_L = -\frac{1}{\alpha-1}$ and $h_R = \frac{1}{\alpha}$. The iso-BAE systems lie on the hyperbolic line $h_R = \frac{h_L}{h_L-1}$. The SU(2) symmetric point is at the origin.}
\label{fig:phase}
\end{figure}

Since all eigenstates are classified into type-I and type-II, one natural question is which class the ground state(s) belong(s) to. It turns out that this depends on the value of $\alpha$ and the ground state degeneracies change at some critical values. To this end, it is more convenient to introduce the boundary magnetic fields
\begin{align}
h_L=-\frac{1}{\alpha-1},\qquad h_R=\frac{1}{\alpha}
\end{align}
Along the iso-BAE flow, they satisfy
\begin{align}
\frac{1}{h_L}+\frac{1}{h_R}=1\,,
\end{align}
which is a hyperbola in the $h_L$-$h_R$ plane, as is shown in figure~\ref{fig:phase}.
The right branch (colored in red) of the hyperbola correspond to $h_R>1$ or equivalently $0<\alpha<1$. The left branch (colored in blue) of the hyperbola correspond to $h_R<1$, which is further divided into $0<h_R<1$ and $h_R<0$, corresponding to $\alpha>1$ and $\alpha<0$ respectively. A quantum phase transition occur at the point $\alpha=0$ and $\alpha=1$. We discuss the cases for even and odd $L$ separately.\\

\begin{figure}
    \centering
\includegraphics[scale=0.5]{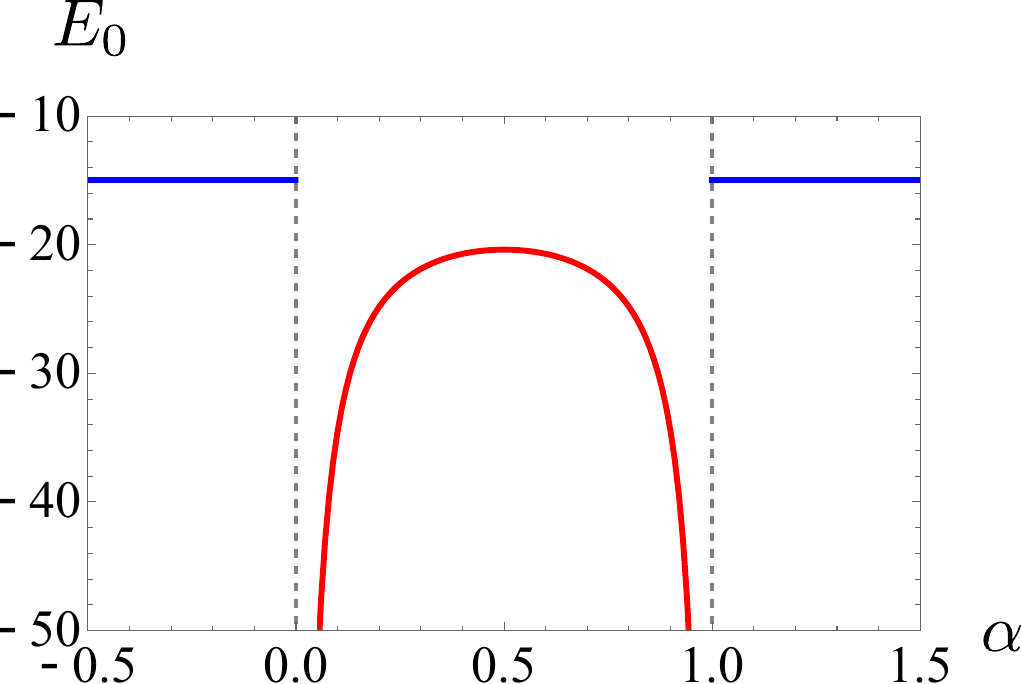}
    \caption{The ground state energy as a function of $\alpha$ for $L=6$. When $\alpha\leq 0$ or $\alpha \geq 1$, the unique ground state is a type-I state with the same ground state energy as the SU(2) point. While $0< \alpha<1$, there are doubly degenerate ground states that are type-II, whose eigenvalues clearly depend on the value of $\alpha$.}    
    \label{fig:groundstateenergy}
\end{figure}

\begin{itemize}
    \item Even $L$: For $0 \leq \alpha \leq 1$, the ground state is a type-II state and is doubly degenerate. The ground state energy varies with $\alpha$. For $\alpha > 1 \cup \alpha < 0$, the ground state is a type-I eigenstate. It is unique and the ground state energy does not vary with $\alpha$. Therefore, when varying $\alpha$ from $-\infty$ to $+\infty$, there will be two phase transition points, corresponding to $\alpha=0$ and $\alpha=1$ where the ground state degeneracies change.
    \item Odd $L$: For $0 \leq \alpha \leq 1 $, the ground states is a type-II. However, it is not doubly degenerate as for even $L$. This is due to the fact that the ground state is in the $(L+1)/2$-magnon sector where $\mathcal{Q}_{\alpha}$ becomes a projector and acts trivially on the type-II state. The energy of the state flows with $\alpha$. For $\alpha > 1\cup\alpha < 0$, the ground state is a type-I, which is unique and does not change with $\alpha$. Therefore in this case the ground state degeneracy do not change, but the type of the ground state is changed at $\alpha=0$ and $\alpha=1$.
\end{itemize}
The ground state energy as a function of $\alpha$ is plotted in Fig.~\ref{fig:groundstateenergy}. As we can see clearly, the type of the ground state changes at $\alpha=0$ and $\alpha=1$. It may seem confusing that the ground state energy jumps from finite value to $-\infty$ when $\alpha$ changes from $1^+$ to $1^-$ (and $0^-$ to $0^+$). We explained in Sec.~\ref{app:alpha1limit} that in the $\alpha \to 1^+$ (and $\alpha \to 0^-$) limit, the ground state energy is the same as the effective Hamiltonian \eqref{eq:Heff} with $(L-1)$ sites, which coincide with the ground state energy of the iso-BAE system with $\alpha < 0$ or $\alpha>1$, \emph{i.e.} the ground state energy of type-I states. However, when we take the other limit $\alpha \to 0^+$ (and $\alpha \to 1^-$), one of the boundary spins is fixed to be $|\downarrow \rangle$, where the ground state energy goes to $-\infty$, as shown in Fig. \ref{fig:groundstateenergy} for instance.

Let us further comment that in \cite{Kattel:2023mfu}, the authors have studied the ground state degeneracies for generic $\alpha$, $\beta$ in the thermodynamic limit. Our study is different from \cite{Kattel:2023mfu} in two aspects. Firstly we focus on the iso-BAE flow, which are the models on the two hyperbolas. Second, our result is valid at any finite length of the spin chain.

\end{appendix}

\end{document}